\documentclass[manuscript]{aastex62}
\usepackage{epstopdf}
\usepackage{graphics}
\usepackage{graphicx}

\begin{document}

\title{FUV to FIR SED modelling of NGC 205}

\author{Denis A. Leahy, Jakob Hansen}
\affil{Department of Physics and Astronomy, University of Calgary, Calgary, AB T2N 1N4, Canada}

\author{Andrew M. Hopkins}
\affil{School of Mathematical and Physical Sciences, Macquarie University, Sydney, NSW 2109, Australia}



\begin{abstract}
New far ultraviolet imaging of the galaxy NGC 205 is presented, which shows the emission is significantly offset ($\sim5^{\prime\prime}$ NW) from the optical and infrared centers of the galaxy.
Spectral energy distribution (SED) modelling is applied to investigate the spatial dependence of the star formation history (SFH) of NGC 205, using data from far ultraviolet to far infrared.
The SED model includes young and old stellar populations, gas emission, dust emission and dust absorption. 
The old stellar population has a total mass of $1.1\times10^8$ M$_{\odot}$ whereas the young population has a much smaller total mass of 3200 M$_{\odot}$.
The best forms of SFH for old and young stars are found to be exponentially declining bursts with start times $t_0$ yr ago and e-folding times $\tau$ yr.
The old stellar population has uniform $t_0$=9.5 Gyr, with $\tau$ decreasing with radius from 1 Gyr to 500 Myr.
The young stellar population has $t_0$=900 Myr and $\tau$=800 Myr, both uniform across NGC 205. 
The young and old stellar mass surface densities are exponential in radius with scale lengths of 40 and 110 pc, respectively. 
The dust heating has a $\sim$40\% contribution from young stars and $\sim$60\% from old stars.
\end{abstract}

\section{Introduction} \label{sec:intro}
NGC 205 is a dwarf galaxy with distance of 0.809 Mpc \citep{2014ApJ...789..147W}, which is a satellite of M31. 
Local group dwarf galaxies are of special interest because they are close enough for detailed studies that can shed light on their structure and star formation history (SFH), thus on how dwarf galaxies are formed.

\cite{2007MNRAS.382.1552L} carried out a population synthesis analysis of the optical spectrum of NGC 205 with 3 age bins, and found $\simeq$50\% of the stellar mass in $2\times10^8$ to  $6\times10^8$ yr, $\simeq$20\% in age bin $6\times10^8$ to  $2\times 10^9$ yr and $\simeq$30\% in age bin $>2\times 10^9$ yr. 
\cite{2005AJ....129.2217B} use color-magnitude-diagram analysis of Hubble Space Telescope (HST) observations of resolved stars and find SFH with a range of ages, with most recent $\lesssim 3\times10^8$ yr ago.
These contrast with the results of \cite{2005AJ....130.2087D}, in which NIR photometry of the brightest AGB stars yields ages between 0.1 and 1 Gyr, and those of \cite{1998ApJ...499..209W} which finds star formation from $5\times10^8$ yr to a few million yr ago.
\cite{1995ApJ...438..680B} carried out HST far-ultraviolet (FUV) observations of NGC 205 and detected 78 stars.  
The optical and FUV CMDs were modeled by an old stellar population ($\sim10$ Gyr) plus recent star formation at a constant level from $5\times10^8$ yr ago to present.
The metal rich stars in NGC 205 have metallicity [Fe/H]$>-0.7$ dex ($>0.2~Z_{\odot}$) while the old stars have  [Fe/H]$>-1.06$ dex ($>0.087~Z_{\odot}$) \citep{2005AJ....129.2217B}.
Thus with differing results on ages and lower limits on metallicity, the SFH of NGC 205 is currently not well known. 

The dust and gas content of NGC 205 are also of interest. 
The H I data are consistent with rotation \citep{1983ApJ...275..549J}, but the stellar body does not appear to rotate \citep{1991A&A...246..349B}, which suggests an external origin for the gas.
The dust and gas in NGC 205 are concentrated in individual molecular clouds \citep{1997ApJ...476..127Y}. 
\cite{2004AAS...20514108M} mapped the dust in NGC 205 using Spitzer.
There is less gas in the central region of NGC 205 than predicted by galactic models, with one explanation that a large number of supernovae expelled gas from the central regions \citep{2012MNRAS.423.2359D}.
\cite{2016MNRAS.459.3900D} compares the content and origin of the interstellar dust in the dwarf spheroidal galaxies NGC 147, NGC 185, and NGC 205.  
For NGC 185 and NGC 205, the observed dust content is a factor 10 higher than that estimated from dust production by AGBs and SNe, whereas NGC 147 has no detected dust, consistent with its evolved stellar population.

In this study, we apply spectral energy distribution (SED) modeling, using the SED fitting code CIGALE \citep{2022ApJ...927..192Y}, to far-ultraviolet to far-infrared observations covering NGC 205.
The SED modeling yields the spatial dependence of stellar, dust, and gas properties. 

The data processing is described in Section~\ref{sec:obs}.
The modeling using CIGALE is described in Section~\ref{sec:meth}.
The results from the SED modeling are given in Section~\ref{sec:results}. 
The discussion of results and comparison with previous work are in Section~\ref{sec:disc}. 
A brief summary is given in Section~\ref{sec:conc}

\section{Observations and Data Processing}  \label{sec:obs}

The data for NGC 205 includes new images, obtained for this study, in three bands in the FUV obtained with the UVIT instrument \citep{2017AJ....154..128T,2020AJ....159..158T} 
on the ASTROSAT satellite \citep{2014SPIE.9144E..1SS}. 
UVIT has a 28 arcminute field of view, a spatial resolution of $\simeq1^{\prime\prime}$, and a variety of FUV and NUV filters. 
The NUV detector failed prior to the observations of NGC 205, thus we obtained images with UVIT in the filters: F148W (120-180 nm bandwidth), F169M (145-175 nm) and F172M (160-185 nm), described in \citep{2017AJ....154..128T}. 

\begin{figure*}[htbp]
    \centering
\includegraphics[scale=0.266]{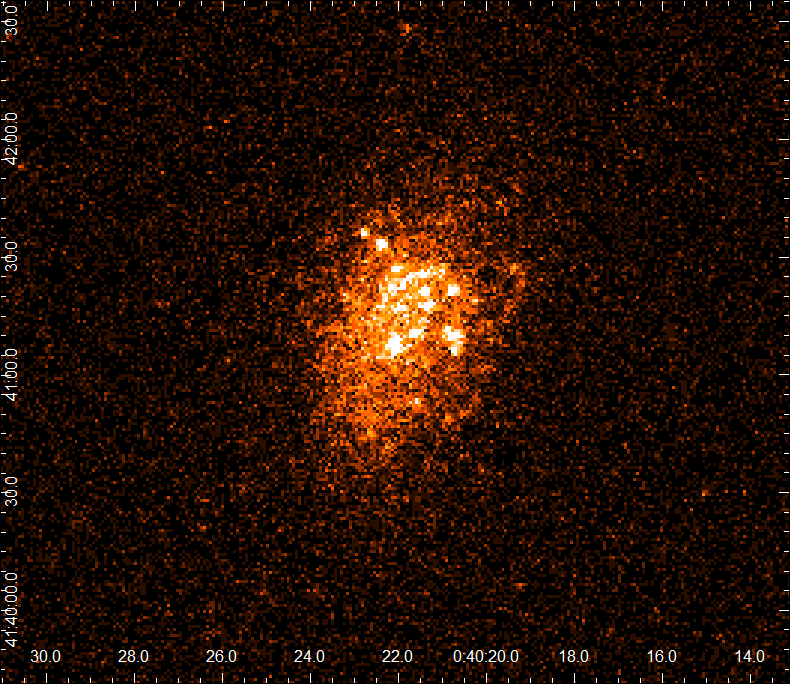}
\includegraphics[scale=0.28]{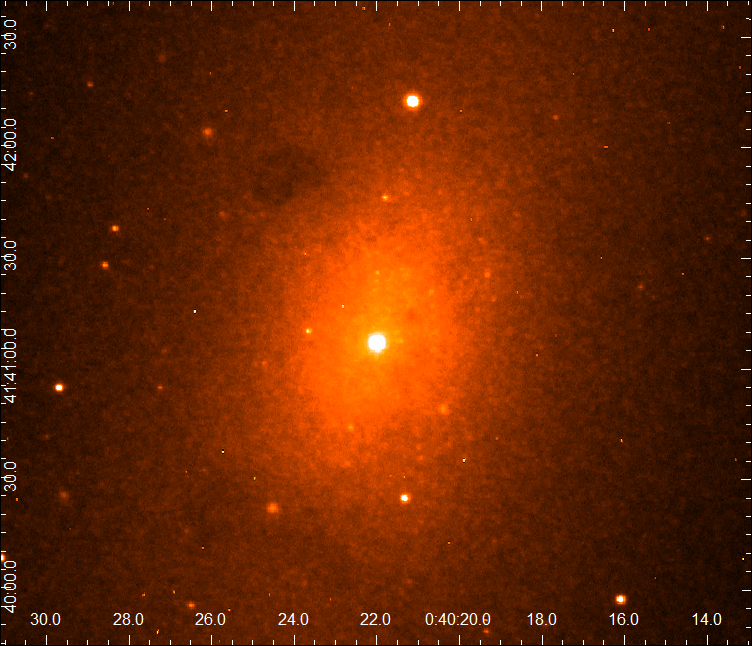}
\includegraphics[scale=0.283]{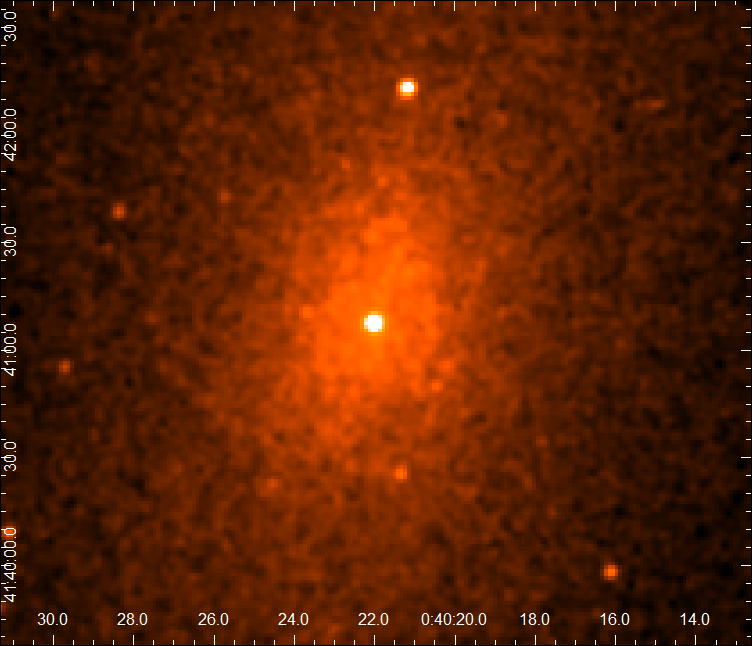}
\includegraphics[scale=0.28]{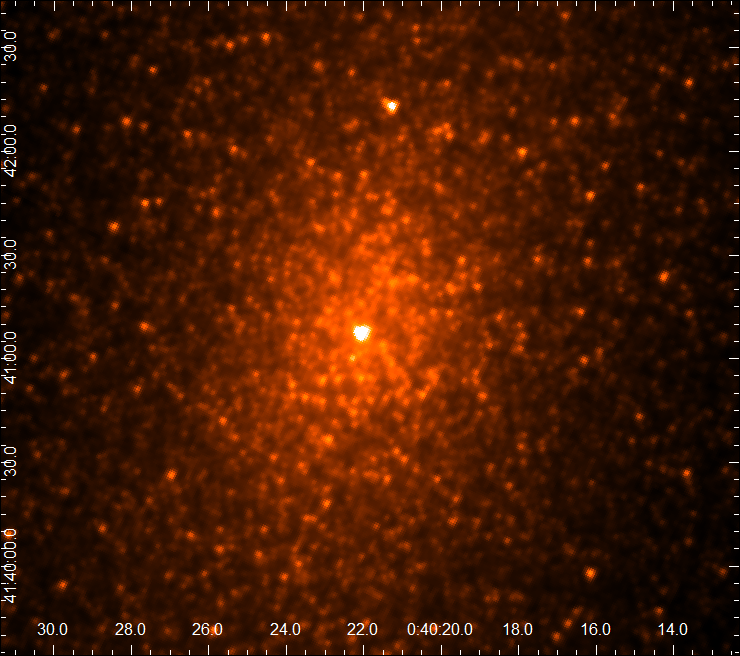}
\includegraphics[scale=0.28]{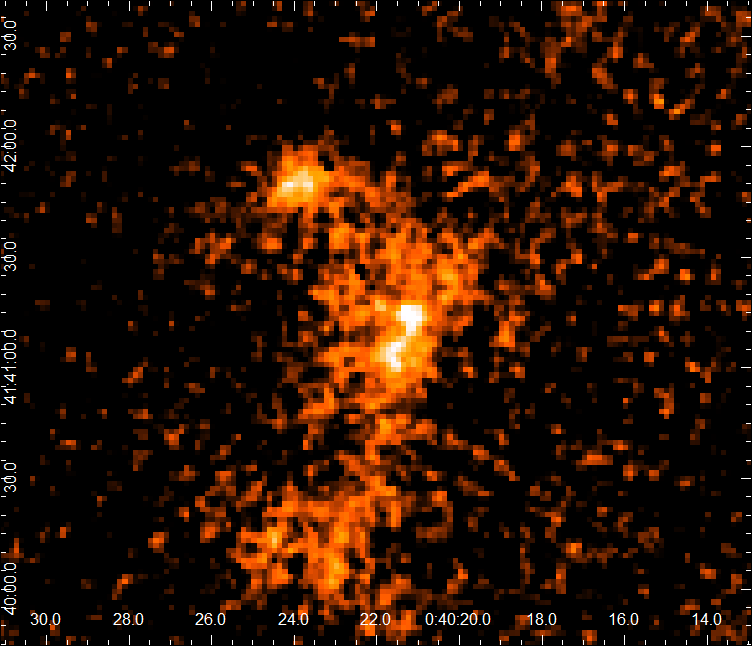}
    \caption{Images of NGC 205 in 5 bands , one for each Telescope listed in Table~\ref{tab:filtertable} (top left to bottom right): 
    (a) UVIT F148W image; (b) JKT V band image;  (c) 2MASS J image; (d) Spitzer 3.5 $\mu$ image; (e) Herschel 70 $\mu$ image.
    \label{fig:bands}
    }    
\end{figure*}

The UVIT images were supplemented with images in 10 different wavebands from observations of NGC 205 in far-infrared (FIR), near-infrared (NIR), and optical domains which were publicly available. 
Table ~\ref{tab:filtertable} lists the filters, their central wavelengths and the telescope (or survey) which carried out the observations.
The last three wavebands (Herschel 250$\mu$, 350$\mu$ and 500$\mu$) were not used in this analysis because their angular resolution was not good enough.
The images of NGC 205 in one selected band from each telescope are shown in Figure ~\ref{fig:bands}.
A few important features of these images are as follows:
the center of the galaxy is seen as a bright spot in the V band, J band and 3.5$\mu$ images; 
the FUV emission from NGC is significantly offset from the center of NGC 205 (by $\sim5^{\prime\prime}$ to the NW) but has a bright spot coincident with the center; 
and the 70$\mu$ image (dust emission) has irregular structures NNE and SSE of the center of NGC 205.

\begin{table}[!ht]
\begin{center}
    \caption{Filters, central wavelengths ($\lambda_c$) and telescopes (or surveys).}
    \label{tab:filtertable}
    \begin{tabular}{|l|l|l|l|}
    \hline
        Filter$^{a}$ & $\lambda_c$ (m) & Telescope$^{b}$ & Reference \\ 
        \hline
        F148W & 1.48$\times10^{-7}$ & UVIT & this work \\ 
        F169M & 1.61$\times10^{-7}$ & UVIT& this work  \\ 
        F172M & 1.72$\times10^{-7}$ & UVIT& this work  \\ 
        V & 5.50$\times10^{-7}$ & JKT & \cite{2007MNRAS.382.1552L}\\ 
        R & 7.00$\times10^{-7}$ & JKT & \cite{2007MNRAS.382.1552L}\\ 
        J & 1.20$\times10^{-6}$ & 2MASS & \cite{2003AJ....125..525J}\\ 
        H & 1.60$\times10^{-6}$ & 2MASS & \cite{2003AJ....125..525J}\\ 
        K & 2.20$\times10^{-6}$ & 2MASS & \cite{2003AJ....125..525J} \\ 
 IRAC\_I1 & 3.55$\times10^{-6}$ & Spitzer & \cite{2004AA...416...41X} \\ 
       IRAC\_I2 & 4.49$\times10^{-6}$ & Spitzer & \cite{2004AA...416...41X} \\ 
  MIPS24 & 2.40$\times10^{-5}$ & Spitzer & \cite{2012MNRAS.423..197B} \\ 
        PACS 70 & 7.00$\times10^{-5}$ & Herschel & \cite{2012MNRAS.419.1833B}\\ 
        PACS 160 & 1.60$\times10^{-4}$ & Herschel &\cite{2012MNRAS.419.1833B} \\ 
        PSW & 2.50$\times10^{-4}$ & Herschel &\cite{2012MNRAS.419.1833B} \\ 
        PMW & 3.50$\times10^{-4}$ & Herschel&\cite{2012MNRAS.419.1833B}  \\ 
        PLW & 5.00$\times10^{-4}$ & Herschel &\cite{2012MNRAS.419.1833B} \\ 
\hline
    \end{tabular}
    \end{center}
\footnotesize
\tablenotetext{a} {For the JKT telescope the filters are Johnson V and Cousins R.}
\tablenotetext{a} {UVIT is the Ultraviolet Imaging Telescope \citep{2017AJ....154..128T}. JKT is the Jacobus Kapteyn Telescope described at https://www.ing.iac.es//astronomy/telescopes/jkt/.
2MASS is the Two Micron All Sky Survey \citep{1997ASSL..210...25S}. 
Spitzer is the Spitzer Space Telescope \citep{2004ApJS..154....1W}. 
Herschel is the Herschel Space Observatory \citep{2010A&A...518L...1P}.}
\end{table}

One of the main goals of the current work is to look for relations between the stellar population properties and the distance from the galactic center of NGC 205.
The center was found by analysing images of NGC 205 in the optical filters ( V and R) which are dominated by stellar emission. 
Those images shared a common position associated with the peak of emission: at ($\alpha$= 00:40:21.99, $\delta$= +41:41:06.9). 

\begin{figure*}[htbp]
    \centering
\includegraphics[scale=1.1]{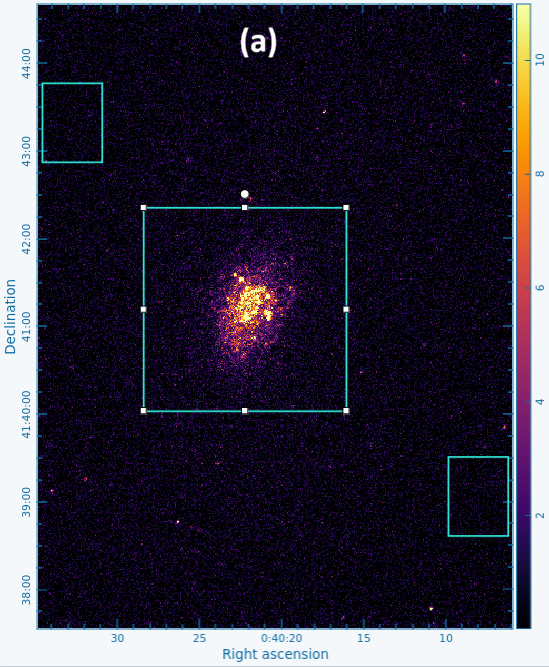} 
\includegraphics[scale=0.485]{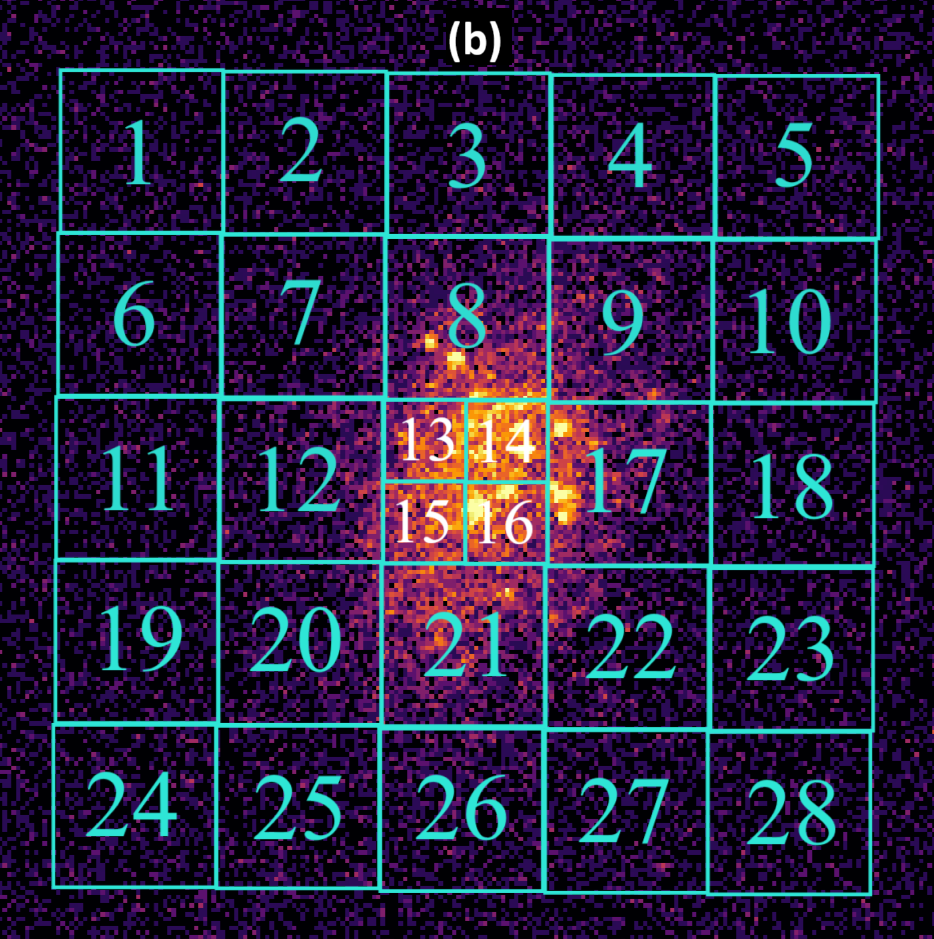}
    \caption{(a) The UVIT F148W (150 nm) image of the NCG 205 area, with area studied (central large box) and the background regions (upper left and lower right boxes). 
    (b) Expanded view of the area studied. 
    The numbered rectangles are the regions analyzed.  
    \label{fig:F148W}
    }    
\end{figure*}

Figure ~\ref{fig:F148W}(a) shows the UVIT F148W image of NGC 205 and its surrounding area.
To model the spatial dependence of NGC 205, it was divided into 28 regions, shown in Figure ~\ref{fig:F148W}(b).
The small regions around the center are $\simeq14^{\prime\prime}$ on a side whereas the large outer regions are $\simeq28^{\prime\prime}$ on a side. 
The photometric data was extracted using CARTA \citep{2021zndo...3377984C}.

Some emission from the outer faint parts of M31, which is located SE of NGC 205, overlaps NGC 205. 
To remove background contributed by M31 and the general sky background, two rectangular background regions were chosen.
They are equidistant from M31, one on either side of NGC 205, and within the field of view of all different waveband images. 
The first background region has NE corner at RA, Dec. ($\alpha,\delta$= 00:40:09.83, +41:38:37.4) and SW corner at ($\alpha,\delta$= 00:40:06.16, +41:39:31.3).
The second has NE corner at ($\alpha,\delta$= 00:40:34.60, +41:42:53.1) and SW corner at ($\alpha,\delta$= 00:40:30.95, +41:43:47.1). 
The background regions have the same area, avoid bright point sources and are shown by the two rectangles in Figure~\ref{fig:F148W}(a).
The background brightnesses (flux densities per sq. arcsec) were extracted for each filter. 
The area-scaled background flux densities were subtracted from the flux densities of each region to obtain source flux densities corrected for sky and M31 background. 
There is significant flux from NGC 205 (above background) from all regions. 
The NIR to FIR emission from NGC 205 is more extended than the FUV emission which can be seen in Figure~\ref{fig:bands}\footnote{The full set of Herschel FIR images of NGC 205 are shown in \cite{2012MNRAS.423.2359D}.}. 

\subsection{UVIT Data}  \label{sec:uvit}

The UVIT images were processed using the CCDLAB software \citep{2017PASP..129k5002P,2022SPIE12181E..3HL}, with astrometry calibration from \citet{2020PASP..132e4503P}. 
The exposure times are F148W: $7.61\times10^{3}$ s, F169M: $6.39\times10^{3}$ s, and F172M: $7.97\times10^{3}$ s. 
The UVIT detectors are photon counting detectors, with practically no dark current, so that the errors are dominated by Poisson statistics\footnote{The majority of image pixels have 0 counts with Poisson error of 1; a smaller number of pixels have 1 count with Poisson error of 1; brighter areas with N counts have a Poisson error of $\simeq \sqrt{N}$.}. 
The UVIT images are in counts/s, which are converted to flux in erg/cm$^2$/s using the conversion factors for each UVIT filter band in \cite{2017AJ....154..128T} and \cite{2020AJ....159..158T}.

\subsection{Optical and Infrared Data for NGC 205}  \label{sec:optir}

Images were obtained from the public NASA/IPAC archives: for the Jacobus Kapteyn Telescope (JKT), 2MASS survey, the Spitzer IRAC and MIPS instruments, and the Herschel PACS instrument. 
Table \ref{tab:filtertable} lists the images' center wavelengths and telescopes.

CIGALE uses input photometry in units of mJy.
All of the image units were converted to mJy/sr then the flux densities in mJy for each region were obtained by summing over the pixels in each region then multiplying by the pixel size in sr.  
The 13-band flux densities for three sample regions, one outer region, one intermediate region, and one central region, are given in Table~\ref{tab:phot}: 
$S$ and $Err$ are the flux density and its error, both in mJy. 
Region 13 is brighter than regions 1 and 7, but is one quarter the area resulting in lower flux density than region 7.

\begin{table}[!ht]
\begin{center}
    \caption{Photometry for select regions}
    \label{tab:phot}
    \begin{tabular}{|l|l|l|l|l|}
    \hline
 & Region 1 & Region 7 & Region 13 \\
Filter & S(Err) & S(Err) & S(Err) \\
        \hline
        F148W & 0.013(0.002) & 0.046(0.007) & 0.17(0.02)\\ 
        F169M & 0.014(0.002) & 0.064(0.006) &0.19(0.03)\\ 
        F172M & 0.027(0.004) & 0.068(0.010)  & 0.24(0.03)\\ 
        V & 3.5(0.8) & 7.0(1.5) &4.1(0.9)\\ 
        R & 4.8(1.8) & 9.9(3.5) &5.4(1.9)\\ 
        J & 19.7(2.7) & 37.8(3.9)  &20.5(2.1)\\ 
        H & 20.6(2.6) & 41.2(4.4) & 22.6(2.4)\\ 
        K & 16.2(2.2) & 33.6(3.7) & 18.1(1.9)\\ 
        SPITZER\_I1 & 13.2(2.3)& 26.5(4.2)& 13.6(2.0) \\ 
        SPITZER\_I2 & 8.5(1.7) & 16.7(2.8) &8.4(1.3)\\ 
        MIPS24 & 1.16(0.15) & 9.7(1.4) & 3.8(0.5)\\ 
        PACS 70 & 8.2(2.7)& 110(17) & 37.6 (5.6)\\
        PACS 160 & 11.4(2.7) & 392(56) & 113(16)\\ 
\hline
    \end{tabular}
    \end{center}
\end{table}

\section{Methodology} \label{sec:meth}

The SED from FUV to FIR of a galaxy (here NGC 205) has three main contributions: from the stars (stellar photospheres, mainly continuum radiation in the FUV, optical and NIR bands); from the dust (primarily continuum radiation in NIR and FIR bands); and from emission nebulae (primarily line radiation from hot gas around the stars). 
The stellar emission is subject to attenuation by dust. 

The SED modelling was carried out with the CIGALE\footnote{Because NGC 205 has no AGN emission, CIGALE is used without invoking its AGN module.} software package \citep{2022ApJ...927..192Y}.
CIGALE is used to compute a grid of SEDs based on user input parameters.
Then CIGALE calculates the $\chi^2$ fit of each SED (with free normalizations) to the input multi-band flux densities (in our case, 13 bands). 
The un-normalized likelihoods, $exp(-\chi^2/2)$, are calculated for each model (labeled as model $j=1...N$, with $N=$ total number of models), then the likelihood normalization factor is obtained as $\frac{1}{\sum_{j=1,N} exp(-\chi_j^2/2)}$.
The Bayes-estimated best parameters parameter \citep{2022ApJ...927..192Y} is the likelihood-weighted average and the Bayes-estimated error is the square-root of the likelihood weighted standard deviation (SD) for each parameter.

Fourteen input parameters, not including 3 SED normalization parameters, describe the stellar emission, the nebular emission, the dust emission and the dust absorption. 
The 3 SED normalization parameters are the stellar masses of old and of young populations and the dust luminosity.
These are determined by fitting the SED to the data and thus are outputs of the fitting process. 
The 14 input parameters and 3 normalizations (converted to mass per unit area or dust luminosity per unit area)  are listed in column 1 of Table~\ref{tab:meanSD}.

Typically 3 values were specified for each of the 14 parameters, resulting in large SED model grid.
It was not possible to centre the initial parameter grid  on the (as yet unknown) Bayes-estimated best parameters nor was the grid suitably spaced around them\footnote{Details of the iterative process using CIGALE are given in Section 3 of \citet{2024AJ....167..211L}}.
Thus the grid of parameters was adjusted based on previous runs for that region.
Then a new grid of SED models was calculated using CIGALE allowing improved Bayes-estimated best parameters and Bayes-estimated errors to be calculated. 
The final input grid of parameters was different for each of the 27 regions. 
The aim was to have the final grid to consist of the Bayes-estimated best value and that plus and minus the 1 and 2 times error for each parameter. 

The different functional forms of star formation were compared for NGC 205. 
Two exponentially declining populations, the ``main" (or old) population and a ``burst" (or young) population fit the data better than alternate forms.
The star formation rate of each group follows $e^{-(t-t_0)/\tau}$ for $t>t_0$ (0 for $t<t_0$). 
The age is the time of the first star formation for that age group (measured from the present time), while $\tau$ is the rate at which stellar formation decreases for that group.
Each group has its three parameters: age ($t_0$), e-folding time ($\tau$) and normalization. 
The metallicity of the stellar populations is a parameter common to both groups. 

\section{Results} \label{sec:results}

\begin{figure*}
     \centering
\includegraphics[scale=0.7]{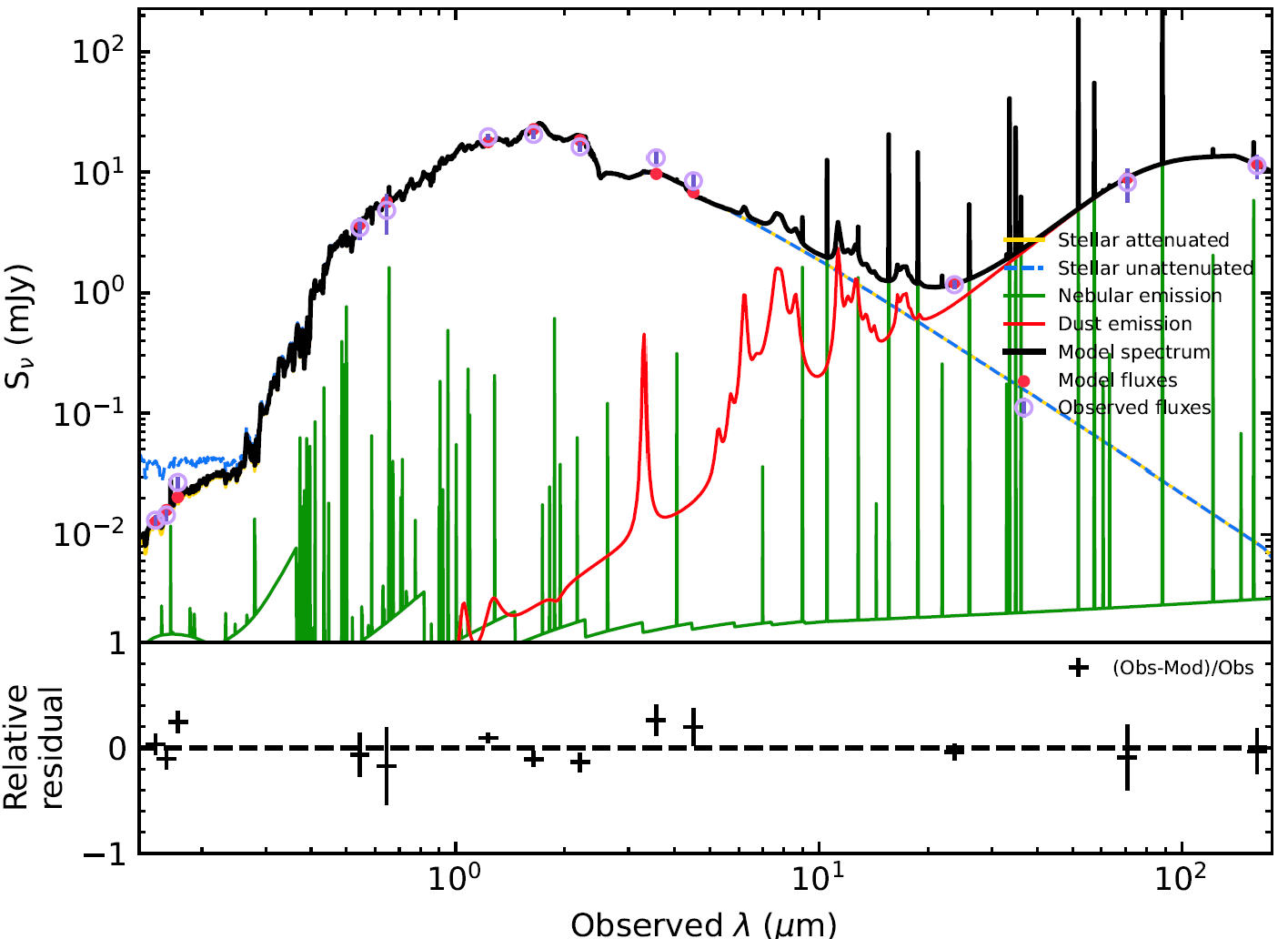}  
    	    \caption{Best-fit model and residual differences between model and data for Region 1.  The legend indicates symbols for input data, model points, and the contributions from the different components of the model flux density. The figures for the Regions 2 to 28 are in the Appendix.  \label{fig:reg1fit}}
\end{figure*}

The SED models fit well the photometry for the different regions of NGC 205. 
Figure~\ref{fig:reg1fit} shows the best-fit model and data for Region 1.
Figure~\ref{SEDfits} in (new) Appendix 1 shows the fits for the other regions (Regions 2 to 28).
Table~\ref{tab:meanSD} gives the mean values and standard deviations for the 28 regions of the different parameters . 
The normalization parameters, which scale with the area of a region, were divided by the area of that region to obtain the surface densities in Table~\ref{tab:meanSD}. 



\begin{table*}[h!]
\begin{center}
	\caption{Means, standard deviations and errors of the SED parameters}
\label{tab:meanSD}
\begin{tabular}{|l|l|l|l|l|}
\hline
Parameter$^{a,b}$&  Mean&  SD & Error Mean&  Error SD\\
\hline 
\hline
Without distance trend && &&\\     
\hline
$t_{main}$ SFH.age\_main (Myr) &  9490&  210& 96& 30\\
$t_{burst}$ SFH.age\_burst (Myr) &  930&  170& 46& 16 \\
$\tau_{burst}$ SFH.tau\_burst (Myr) & 760&  90&38 & 7\\
$Z_{stars}$ stellar\_metallicity$^{c}$&0.050&4$\times10^{-4}$&0.003 & 0.002\\
$f_{E(B-V)}$ attenuation.E\_BV\_factor &  0.385& 0.155 & 0.036& 0.011  \\    
$E(B-V)_{lines}$ attenuation.E\_BV\_lines &  0.136&  0.074 & 0.017 & 0.013 \\
 $\beta$ attenuation.powerlaw\_slope &  $-$1.41&  0.59&0.12 &0.05 \\
$f_{dust}$ nebular.f\_dust &  0.30&  0.04& 0.020&0.001  \\
$Z_{gas}$ nebular.zgas$^{c}$ &  0.0204&2$\times10^{-4}$ & 0.002&0.001 \\
$q_{pah}$ dust.qpah &  5.5&  1.6& 0.6&  0.1\\
$U_{min}$ dust.umin &  1.38&  0.76 &0.28 &0.12 \\
$\alpha_{dust}$ dust.alpha  &  2.73&  0.14&0.11 & 0.02 \\
$\gamma_{dust}$ dust.gamma  &  0.36&  0.12 &0.04 &0.01 \\
\hline  
\hline
With distance trend && & & \\     
\hline
$\sigma_{main}$ mstar\_old/area (M$_{\odot}$/sr)$^d$ &  $2.9\times10^{14}$ &  $1.5\times10^{14}$ & $1.4\times10^{13}$ &  $7.5\times10^{12}$  \\
$\sigma_{burst}$ mstar\_young/area  (M$_{\odot}$/sr)$^d$ &  $1.2\times10^{10}$ &  $1.8\times10^{10}$ & $8.5\times10^{8}$ &  $9.2\times10^{8}$ \\
$\tau_{main}$ SFH.tau\_main  (Myr) &  660&  220&44 &12  \\
$I_{dust}$ dust.luminosity/area (erg/s/sr)$^d$ &  $3.7\times10^{39}$ & $3.0 \times10^{39}$ & $2.6\times10^{38}$ & $1.9\times10^{38}$  \\\hline
\hline
Other quantities &&&& \\ 
\hline
$\chi^2_r$ best\_reduced\_chi\_square  &  0.58&  0.15 & & \\
\hline
\end{tabular}
\end{center}
\footnotesize
\tablenotetext{a}{The symbol and the CIGALE parameter name are given, then units in brackets. Parameters without specified units are dimensionless. Definitions for each variable are found in Section 3 of \cite{2024AJ....167..211L}, with details in \cite{2022ApJ...927..192Y}. SD stands for standard deviation.}
\tablenotetext{b}{The SED input parameters are the first 13 plus $\tau_{main}$.}
\tablenotetext{c}{The metallicity is in absolute units, ie ratio of abundance of metals to that of hydrogen. To convert to relative metallicity, divide by the metallicity of the Sun $Z_{\odot}=0.02$.}
\tablenotetext{d}{At the distance of NGC 205, 1 pc$^2$ area is $1.528\times10^{-10}$ sr.}
\end{table*}

To determine whether any of the parameters have a distance trend, minimum $\chi^2$ fitting was utilized.
The functions tested were: 
\begin{eqnarray}
f_{C}(d)=a_0 &~~~~~~ \rm{constant} \nonumber \\
f_{lin}(d)=a_1~ d +a_0&~~~~~~ \rm{linear} \nonumber \\
f_{exp}(d)=a_1~e^{-d/a_2}+a_0&~~~~~~  \rm{exponential} \nonumber \\
f_{G}(d)=a_1~ e^\frac{-(d-a_3)^2}{2~ a_2^2}+a_0&~~~~~~  \rm{Gaussian} \nonumber \\
\end{eqnarray}
where $d$ is the distance from the centre of NGC 205. 

The exponential was the best fit for the four SED parameters listed in Table~\ref{tab:var}. 
For the other parameters, the constant function was as good a fit as the other functions.

\begin{table*}[h!]
\begin{center}
\caption{Best-fit exponential functions ($a_1~e^{-D/a_2}+a_0$) and their 1$\sigma$ errors for SED parameters with distance trend.}
    \centering
    \begin{tabular}{|l|l|l|l|}
    \hline
         & $a_0$ (error) & $a_1$(error) & $a_2$ (error) (pc) \\ 
        \hline
        $\sigma_{main}$ (M$_\odot$/sr) & 1.0$(\pm0.1)\times10^{14}$  & 6.8$(\pm0.4)\times10^{14}$ & 113($\pm3$)\\ 
        $\sigma_{burst}$ (M$_\odot$/sr) & 8$(\pm4)\times10^{8}$ & 1.1$(\pm0.2)\times10^{11}$& 38($\pm$5)\\ 
        $\tau_{main}$ (Myr) & 490($\pm43$) & 625($\pm$66)& 93($\pm$15)\\ 
        $I_{dust}$ (erg/s/sr) & 1.4$(\pm0.2)\times10^{39}$&1.5$(\pm0.3)\times10^{40}$ & 50($\pm$9)\\ 
        \hline
    \end{tabular}
    \label{tab:var}
\end{center}
\end{table*}

\section{Discussion} \label{sec:disc}

There have not been many measurements of the structure or SFH of dwarf elliptical galaxies like NGC 205 previously.
One of the main goals here is to provide basic measurements of SFH and structure so that they can be understood by comparison with other dwarf elliptical galaxies.

\subsection{Parameters without distance dependence}\label{sec:noD}

Most of the SED parameters did not show any distance dependence (all but the four listed in Table~\ref{tab:meanSD}).
The parameters without distance dependence can be separated into two sets: those consistent with a constant or those that vary without a clear pattern with distance. 
The criterion is used that if the standard deviation seen in Table~\ref{tab:meanSD} is less than $10\%$ of the mean, the parameter is effectively constant throughout NGC 205
The parameters that stay constant are: $t_{main}$, $Z_{stars}$, $Z_{gas}$ and $\alpha_{dust}$.  $\tau_{burst}$, and $Z_{stars}$. 
Parameters with standard deviations slightly greater (between $10\%$ and $20\%$ of the mean) are $t_{burst}$ and $f_{dust}$.
There are 6 parameters that change within NGC 205 but have no systematic pattern with distance. 
They are either dust properties ($\beta$, $q_{pah}$ and $\gamma_{dust}$) or related to the distribution of dust and gas relative to the stars ($E(B-V)_{lines}$, $U_{min}$ and $f_{E(B-V)}$\footnote{$f_{E(B-V)}$ is the ratio of $E(B-V)_{stars}$ to $E(B-V)_{lines}$.}). 

The gas distribution in NGC 205 was discussed in \cite{1997ApJ...476..127Y}, based on HI and CO observations. 
Both distributions are clumpy, with the HI map showing 3 peaks in column density: one near the centre of NGC 205 ($\alpha,\delta$= 0:40:22, +41:10:00), one in the SE ($\alpha,\delta$= 0:40:23, +40:08:00) and one in the NE ($\alpha,\delta$= 0:40:23.5, +41:48:00). 
The peak column densities are largest in the SE ($N_H=3.8\times10^{20}$ cm$^{-2}$), intermediate in the centre ($N_H=2.6\times10^{20}$ cm$^{-2}$) and lowest in the NE ($N_H=2.2\times10^{20}$ cm$^{-2}$). 

\begin{figure}[htbp]
    \centering
\includegraphics[width=1.03\linewidth]{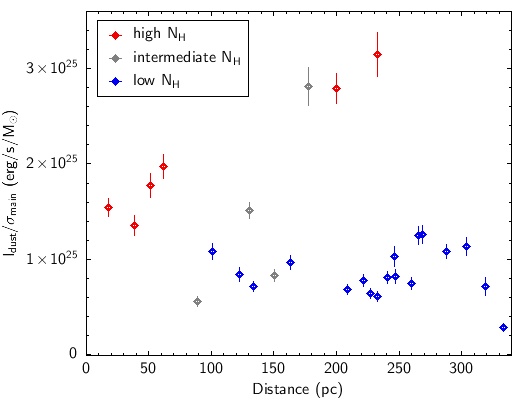}   
\caption{Dust luminosity per mass of old stars for regions with high, intermediate and low $N_H$. \label{fig:dustNH}
}
\end{figure}

The region boundaries (Figure~\ref{fig:F148W}a) were overlaid on the HI contours and the regions were separated into 3 groups: group 1 with $N_H<8\times10^{19}$ cm$^{-2}$, group 2 with
$8\times10^{19}<N_H<1.8\times10^{20}$ cm$^{-2}$ and group 3 with $N_H>1.8\times10^{20}$ cm$^{-2}$.
There are 6 regions in group 3 (high $N_H$), 4 in group 2 (intermediate $N_H$) and 18 in group 1 (low $N_H$). 
We searched for correlations of SFH parameters with HI column density but only two were found, and these were for derived parameters. 
$I_{dust}$ is higher for high $N_H$ regions than for low $N_H$ regions. 
A similar but stronger correlation is seen in the dust luminosity per mass of old stars, shown in Figure~\ref{fig:dustNH}.
This shows a peak for the regions of the SE HI peak (at $\sim$200 pc distance).
However, the dust luminosity per mass of young stars is not different between the 3 groups.
This result may be caused as follows: the fairly smoothly distributed old stars exist in the high $N_H$ regions yielding a high dust luminosity per mass, but strong winds and radiation from young stars prevent the high gas and dust density very near young stars. 

The CO observations cover only a small part of NGC 205 and gave only a few detections \citep{1997ApJ...476..127Y}.
Those are mostly near the NE HI peak (in region 7) with one detection E of the SE peak (in region 12). 
No clear difference in SED parameters is found for regions 7 and 12 compared to the other regions. 
 
\subsection{Parameters with distance dependence}

The 4 parameters with distance dependence have their 2-D spatial dependencies shown in Tables 5, 6, 7 and 8. 
The parameters are given in a grid pattern than matches the grid of regions which cover NGC 205, shown in Figure~\ref{fig:F148W}(b). 
Thus the Tables preserve the spatial relations between parameters, e.g. the four numbers in the central box in the Tables are for the small central regions 13, 14, 15 and 16.
The surface mass density of old stars ($\sigma_{main}$) increases from the outer regions towards the central regions (13 to 16) and is highest in region 16 which includes the center of NGC 205. 
The ellipticity of NGC 205 is visible with regions 3 and 26 with higher $\sigma_{main}$ than regions 11 and 18. 
The surface mass density of young stars ($\sigma_{burst}$) is not as smoothly distributed as $\sigma_{main}$.
For example there is a secondary peak (about 1/4 the density of the 4 central regions) south of center in region 26. 
The e-folding time of formation of the old population ($\tau_{main}$) is shown in Table 7. 
The largest $\tau_{main}$ values are in the central 4 regions and in region 26. 
Table 8 shows the values of dust surface brightness ($I_{dust}$). 
The highest values are for the 4 central regions, but also NNE (regions 7 and 8) and SSE of the center (regions 25 and 26).
The high $I_{dust}$ values are located where the brightest regions are located in the Herschel 160$\mu$ map, shown in Figure~\ref{fig:bands}. 

\begin{table}[h!]
\begin{center}
\caption{Mass density of main population $\sigma_{main}$ (M$_{\odot}/sr$) vs. region position$^{a}$}
    \centering
    \begin{tabular}{|c|c|c|c|c|}
    \hline
1.2$\times10^{14}$ & 1.9$\times10^{14}$ & 3.1$\times10^{14}$ & 2.5$\times10^{14}$ & 1.8$\times10^{14}$ \\ 
  & & & & \\
 \hline
 1.4$\times10^{14}$ & 2.5$\times10^{14}$ & 3.9$\times10^{14}$ & 3.2$\times10^{14}$ & 1.7$\times10^{14}$ \\  
    & & & &\\
\hline
1.5$\times10^{14}$ & 3.0$\times10^{14}$ & 4.6$\times10^{14}$ 5.0$\times10^{14}$ & 3.8$\times10^{14}$ &1.9$\times10^{14}$ \\
& &  5.6$\times10^{14}$  7.7$\times10^{14}$  & & \\
\hline
 1.7$\times10^{14}$ & 3.2$\times10^{14}$ & 4.2$\times10^{14}$ & 3.2$\times10^{14}$ & 1.7$\times10^{14}$ \\  
  & & & &    \\
      \hline
 1.8$\times10^{14}$ & 2.2$\times10^{14}$ & 2.5$\times10^{14}$ & 2.3$\times10^{14}$ & 1.3$\times10^{14}$ \\  
       & & & & \\
  \hline
    \end{tabular}
    \label{tab:sigmain}
\end{center}
\footnotesize
\tablenotetext{a}{The grid of regions is shown in Figure~\ref{fig:F148W}.}
\end{table}

\begin{table}[h!]
\begin{center}
\caption{Mass density of burst population $\sigma_{burst}$ (M$_{\odot}/sr$) vs. region position}
    \centering
    \begin{tabular}{|c|c|c|c|c|}
    \hline
6.4$\times10^{8}$ & 1.8$\times10^{9}$ & 1.6$\times10^{9}$ & 1.3$\times10^{9}$ & 2.2$\times10^{9}$ \\ 
  & & & & \\
 \hline
  2.7$\times10^{9}$ &  7.8$\times10^{9}$ &  2.1$\times10^{10}$ & 9.5$\times10^{9}$ &  1.7$\times10^{9}$ \\ 
    & & & &\\
    \hline
    2.2$\times10^{9}$ &  7.0$\times10^{9}$&  4.5$\times10^{10}$   6.3$\times10^{10}$ & 1.7$\times10^{10}$ &  1.7$\times10^{9}$ \\ 
  & & 3.8$\times10^{10}$   6.3$\times10^{10}$ & & \\
     \hline
     2.4$\times10^{9}$ &  8.5$\times10^{9}$ &  4.8$\times10^{9}$ & 5.6$\times10^{9}$ &  5.8$\times10^{8}$ \\ 
  & & & &    \\
      \hline
      6.9$\times10^{8}$ & 8.2$\times10^{9}$ &1.4$\times10^{10}$ &  7.4$\times10^{8}$ &  3.3$\times10^{9}$ \\ 
       & & & & \\
  \hline
    \end{tabular}
    \label{tab:sigburst}
\end{center}
\end{table}

\begin{table}[h!]
\begin{center}
\caption{e-folding times of main population $\tau_{main}$ (Myr) vs. region position}
    \centering
    \begin{tabular}{|c|c|c|c|c|}
    \hline
   500 & 480 &  400 &580 & 450 \\ 
  & & & & \\
 \hline
    500 & 650 & 825 &580 & 590 \\ 
  & & & & \\
\hline
    610 & 630 &  900 1300  &827 & 550 \\ 
& &  960 950   & & \\
\hline
    520 & 560 &  500 &560 & 440 \\ 
  & & & &    \\
  \hline
   490 & 720 & 1000 &450 & 740 \\ 
       & & & & \\
  \hline
    \end{tabular}
    \label{tab:taum}
\end{center}
\end{table}

\begin{table}[h!]
\begin{center}
\caption{Dust surface brightness $I_{dust}$ (erg/s/sr) vs. region position}
    \centering
    \begin{tabular}{|c|c|c|c|c|}
    \hline
3.1$\times10^{38}$ & 2.3$\times10^{39}$ & 2.5$\times10^{39}$ & 1.9$\times10^{39}$ & 1.3$\times10^{39}$ \\ 
  & & & & \\
 \hline
1.7$\times10^{39}$ &6.9$\times10^{39}$ & 5.9$\times10^{39}$ & 3.1$\times10^{39}$ & 1.8$\times10^{39}$ \\ 
    & & & &\\
\hline
9.2$\times10^{38}$ &2.5$\times10^{39}$ & 9.0$\times10^{39}$ 8.8$\times10^{39}$ &4.1$\times10^{39}$ & 1.3$\times10^{39}$ \\ 
& & 7.5$\times10^{39}$ 1.2$\times10^{40}$  & & \\
\hline
1.4$\times10^{39}$ & 2.6$\times10^{39}$ & 2.3$\times10^{39}$ & 2.3$\times10^{39}$ & 1.1$\times10^{39}$ \\ 
  & & & &    \\
      \hline
2.0$\times10^{38}$ & 6.8$\times10^{39}$ & 6.9$\times10^{39}$ & 1.8$\times10^{39}$ & 1.4$\times10^{39}$ \\ 
       & & & & \\
  \hline
    \end{tabular}
    \label{tab:Idust}
\end{center}
\end{table}

\begin{figure}[htbp]
    \centering
    \includegraphics[width=1.03\linewidth]{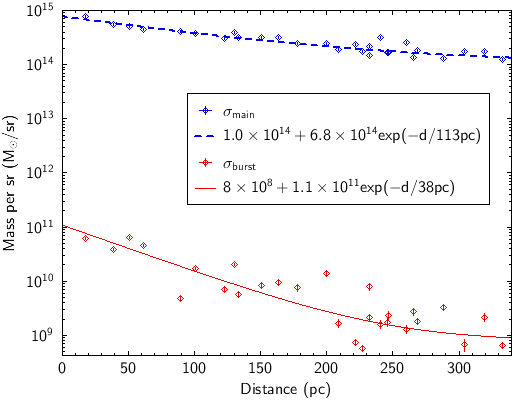} 
\caption{Surface mass density of old stars $\sigma_{main}$ and of young stars  $\sigma_{burst}$ as a function of distance from the centre of NGC 205. 
\label{fig:starold}
}
\end{figure}

The surface mass densities of old and young stellar populations are shown in Figure~\ref{fig:starold}. 
Both stellar components of NGC 205 follow exponential laws in radius, with scale lengths for old and for young stars of 113 pc and 38 pc, respectively. 
The young stars are significantly more concentrated to the center than the old stars.
The scale-lengths for both stellar populations of dwarf galaxies have not been measured previously.

For spiral galaxies, the disks (young stars) are more extended than the bulges (old stars) but for ellipticals the young star distributions are not well measured.
The reason that NGC 205 shows more concentrated young stars could be that it has weaker gravitational field, so that the gas density does not reach the critical density for star formation until close to the center.
The distribution of young stars also depends on the source of gas for star formation.
If the gas comes from existing stars, it tends to be trapped in the center, giving a pattern similar to that for NGC 205.
If the gas is accreted, star formation can peak where the gas reaches the critical density for star formation, which can be anywhere along the path of the accreted gas.

The total mass of old stars for the 27 regions is $1.1\times10^8$ M$_{\odot}$, whereas for young stars it is 3200 M$_{\odot}$.
A possible explanation for the variation in density of young stars (Figure~\ref{fig:starold}) is that stars form in clusters and there is only a small total number of young stars, so that stochastic variations are large.

\begin{figure*}[htbp]
    \centering
     \includegraphics[width=0.495\linewidth]{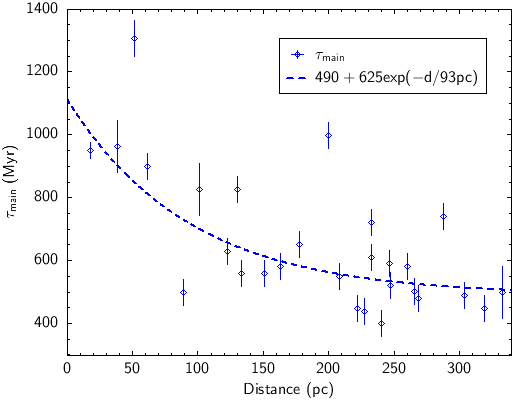} 
    \includegraphics[width=0.495\linewidth]{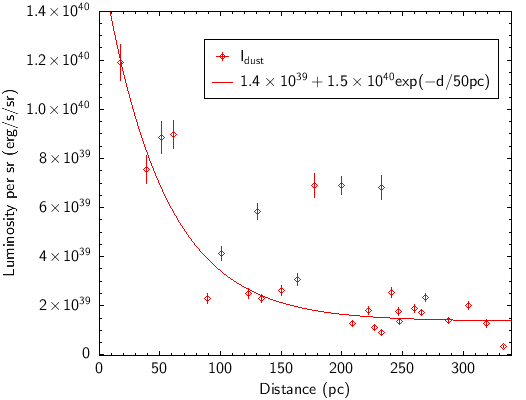} 
\caption{Left: The duration of old star formation ($\tau_{main}$) in Myr vs. distance from the centre of NGC 205.  Right: Dust luminosity per sr $I_{dust}$ (erg/s/sr) vs. distance from the centre of NGC 205. The lines are the best-fit exponential functions.  \label{fig:taum.Id}
}
\end{figure*}

The radial dependence of $\tau_{main}$ is shown in Figure~\ref{fig:taum.Id}a.
It decreases with distance as an exponential with scale length of 93 pc, i.e. similar to the scale length for density of old stars.
With the start time of old star formation in NGC 205 constant at 9.5 Gyr, it is not clear what the cause of the variable $\tau_{main}$, but it may be related to the gas supply at the time of assembly of NGC 205. 

The dust surface brightness $I_{dust}$ is shown in Figure~\ref{fig:taum.Id}b with its exponential fit. 
The scale length is similar to that for young stars, which might indicate that young stars dominate the dust heating.
There are a few data points deviating from the exponential trend with much greater dust surface brightness near 150 to 250 pc. 
These deviations are likely from individual dust clouds inside of NGC 205, which have been observed before  \citep{1997ApJ...476..127Y}, and discussed above in Sec~\ref{sec:noD}.

The surface mass density of old stars is $\sim10^4$ times larger than that of young stars.
But old stars are individually considerably less luminous.
The total luminosity from old stars is $\sim$75 times larger than that from young stars.
Thus both young and old can contribute significantly to dust heating.
To quantify this, one can write:
\begin{equation}
 I_{dust}=   a~ \sigma_{burst} + b ~  \sigma_{main}
\end{equation}
Fitting this to the measured $I_{dust}$ values yields best-fit coefficients of $a=1.51(\pm0.14)\times10^{29}$ erg/s/M$_{\odot}$ and $b=4.97(\pm0.37)\times10^{24}$ erg/s/M$_{\odot}$.
Thus for NGC 205 the dust luminosity per stellar mass is larger for young stars by a factor of $3\times10^4$. 
The total dust heating from young stars is given by $a~M_{burst}$ and from old stars by $b~M_{main}$, which are $4.8\times10^{32}$ erg/s and $5.5\times10^{32}$ erg/s, respectively.
The old star contribution is $\sim$1.5 times larger than that from young stars.
This reflects the balance between the larger total radiation from the numerous old stars and the more effective heating of dust by the UV emission from young stars.

\subsection{Comparisons with other studies}\label{sec:sub_dis_2}

Studies of NGC 205 using individual resolved stars have largely focused on two aspects: star formation rates and metallicity.
Those studies focused on the hot young stars, mainly because the hot young stars were the only stars capable of being resolved. 
The consensus is that NGC 205 currently has active star formation or at least has experienced recent star formation.

\cite{2005AJ....129.2217B} use color-magnitude-diagram analysis of Hubble Space Telescope (HST) observations of resolved stars and find star formation over a range of ages, with most recent $\lesssim$300 Myr ago.
This contrasts with the results of \cite{2005AJ....130.2087D}, in which NIR photometry of the brightest AGB stars yields ages between 100 and 1000 Myr, and those of \cite{1998ApJ...499..209W} which finds star formation from 500 Myr to a few Myr ago.
\cite{1995ApJ...438..680B} carried out HST FUV observations of NGC 205 and studied 78 stars.  
The optical and FUV CMDs were modeled by an old stellar population ($\sim10$ Gyr) plus recent star formation at a constant level from 500 Myr ago to present.

The results here support and refine the above results: the  main star formation event was at 9500 Myr, and there is ongoing star formation from 900 Myr to the present.
From $t_{burst}$, $\tau_{burst}$ and the total mass of young stars, the present star formation rate for the young population is found to be $1.3\times10^{-3}$ M$_{\odot}$/yr.
This is quite high for a low mass elliptical galaxy.

High metallicity is common in elliptical galaxies with old stellar populations, likely due to rapid early star formation cycling through several generations quickly, to build up the metallicity at early times, which results in an old, long-lived, high-metallicity population.
The previous measurement of metallicity for NGC 205 \citep{2005AJ....129.2217B} finds young (300 Myr) high-metallicity stars with $Z>0.2~Z_{\odot}=0.004$ and old stars with $Z>0.087~Z_{\odot}=0.0017$.
The mean metallicity\footnote{Absolute units are used here: ratio of metal abundance to hydrogen abundance.} found from our SED fitting for NGC 205  is 0.05 (Table \ref{tab:meanSD}).
Formally, this result is compatible with the lower limits on metallicity found previously, and we find that the metallicity is uniform across NGC 205. 
However the formal errors for metallicity are likely underestimated in CIGALE, in part because CIGALE has only a course grid of metallicity values and in part because of the limitations of SED fitting in determination of metallicity.
Thus, although there is an indication of high metallicity and possibly uniform metallicity for NGC 205, further observations are needed to confirm this.

\section{Conclusions}  \label{sec:conc}

New FUV images of NGC 205 were obtained with the UVIT instrument on the Astrosat satellite.
The FUV emission is centered $\sim5^{\prime\prime}$ NW of the optical and IR center of NGC 205, but there is compact FUV emission also associated with the optical/IR center.
This study applies SED fitting to observations of NGC 205 across a wide range of wavebands from FUV to FIR, incorporating the new FUV data. 
The FUV data increase the sensitivity of the SED modelling to hot young stars.
The broad-band coverage increases the reliability of the SED modelling. 
A number (28) of regions were separately fit in order to detect spatial variations of the SFH and dust properties of NGC 205.

The age of the older stellar population is 9.5 Gyrs throughout NGC 205.
The young stellar population began to form 900 Myrs ago and is still ongoing, with a current star formation rate of $1.3\times10^{-3}$ M$_{\odot}$/yr.
The total mass of old stars is $\simeq1.1\times10^8$ M$_{\odot}$ and of young stars is $\simeq$3200 M$_{\odot}$.
The old and young stellar components have exponential surface mass densities,
with radial scale lengths of 113 and 38 pc, respectively. 
The dust surface brightness is exponential with radial scale length of 50 pc, which depends both on the stellar heating and gas and dust density distributions.
There is excess dust surface brightness where high column density HI is located.
The old stars contribute $\sim$60\% to dust heating, and the young stars $\sim$40\%.

The SED study of NGC 205 has yielded a number of measurements of basic properties of this dwarf elliptical galaxy.
These should be useful for understanding the nature and evolution of dwarf galaxies as more studies of other dwarf galaxies are carried out.

\paragraph{Acknowledgements}

This publication uses data from the AstroSat mission of the Indian Space Research Institute (ISRO), archived at the Indian Space Science Data centre (ISSDC).
 
This work is based [in part] on observations made with the Spitzer Space Telescope, which was operated by the Jet Propulsion Laboratory, California Institute of Technology under a contract with NASA.

Herschel is an ESA space observatory with science instruments provided by European-led Principal Investigator consortia and with important participation from NASA.
 
This publication makes use of data products from the Two Micron All Sky Survey, which is a joint project of the University of Massachusetts and the Infrared Processing and Analysis Center/California Institute of Technology, funded by the National Aeronautics and Space Administration and the National Science Foundation.

This research has made use of the NASA/IPAC Extragalactic Data base (NED) which is operated by the Jet Propulsion Laboratory, California Institute of Technology, under contract with the National Aeronautics and Space Administration.

We thank the reviewer for comments which significantly improved the manuscript.

\paragraph{Funding Statement}
This project is undertaken with the financial support of the Canadian Space Agency.

\paragraph{Competing Interests}
None.


\paragraph{Ethical Standards}
The research meets all ethical guidelines, including adherence to the legal requirements of the study country.

\paragraph{Author Contributions}
Conceptualisation: D.L. Methodology: D.L. Data and modelling: J.H. Data visualisation: D.L., J.H., A.H. Writing: D.L., A.H., J.H. All authors approved the final submitted draft.

\printendnotes

\bibliography{main}

\appendix
\section{SED fits for the NGC 205 Regions}

The SED fit for Region 1 was shown in Figure~\ref{fig:reg1fit} in the main text. Those for the remaining 27 regions (2 to 28) are given here in Figures 7 to 11. 

\begin{figure*}
     \centering
\includegraphics[scale=0.48]{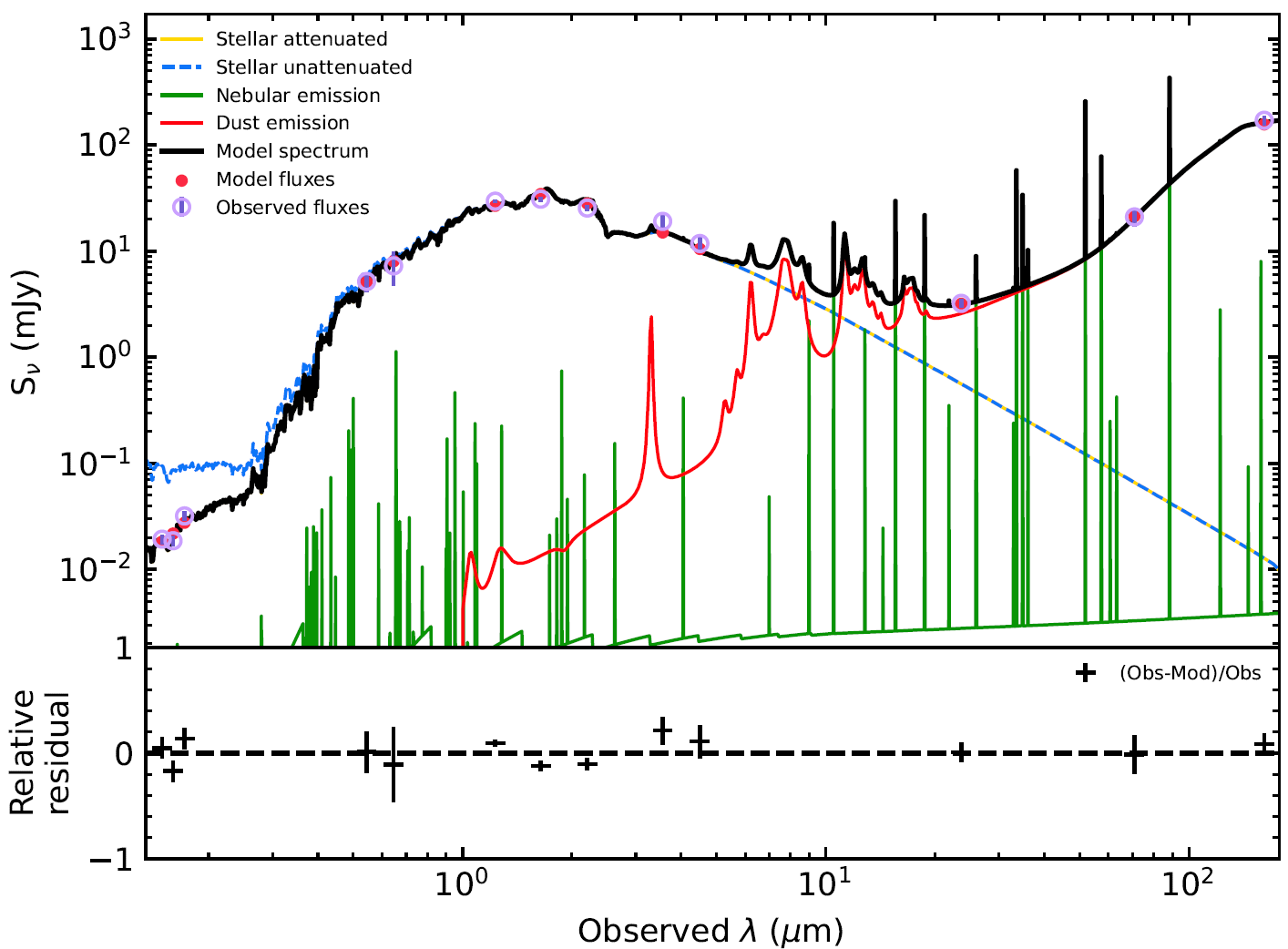}   \includegraphics[scale=0.48]{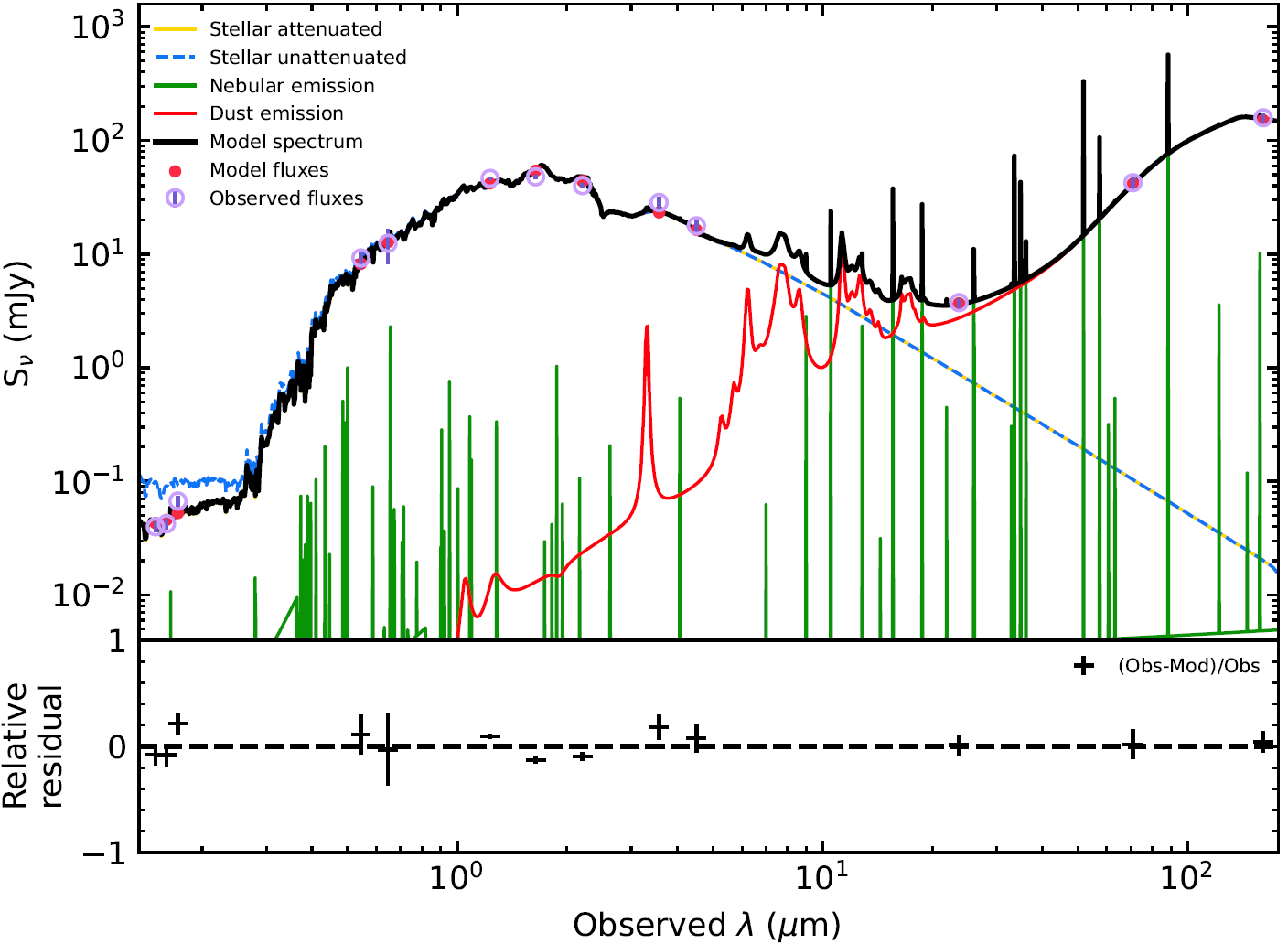} 
\includegraphics[scale=0.48]{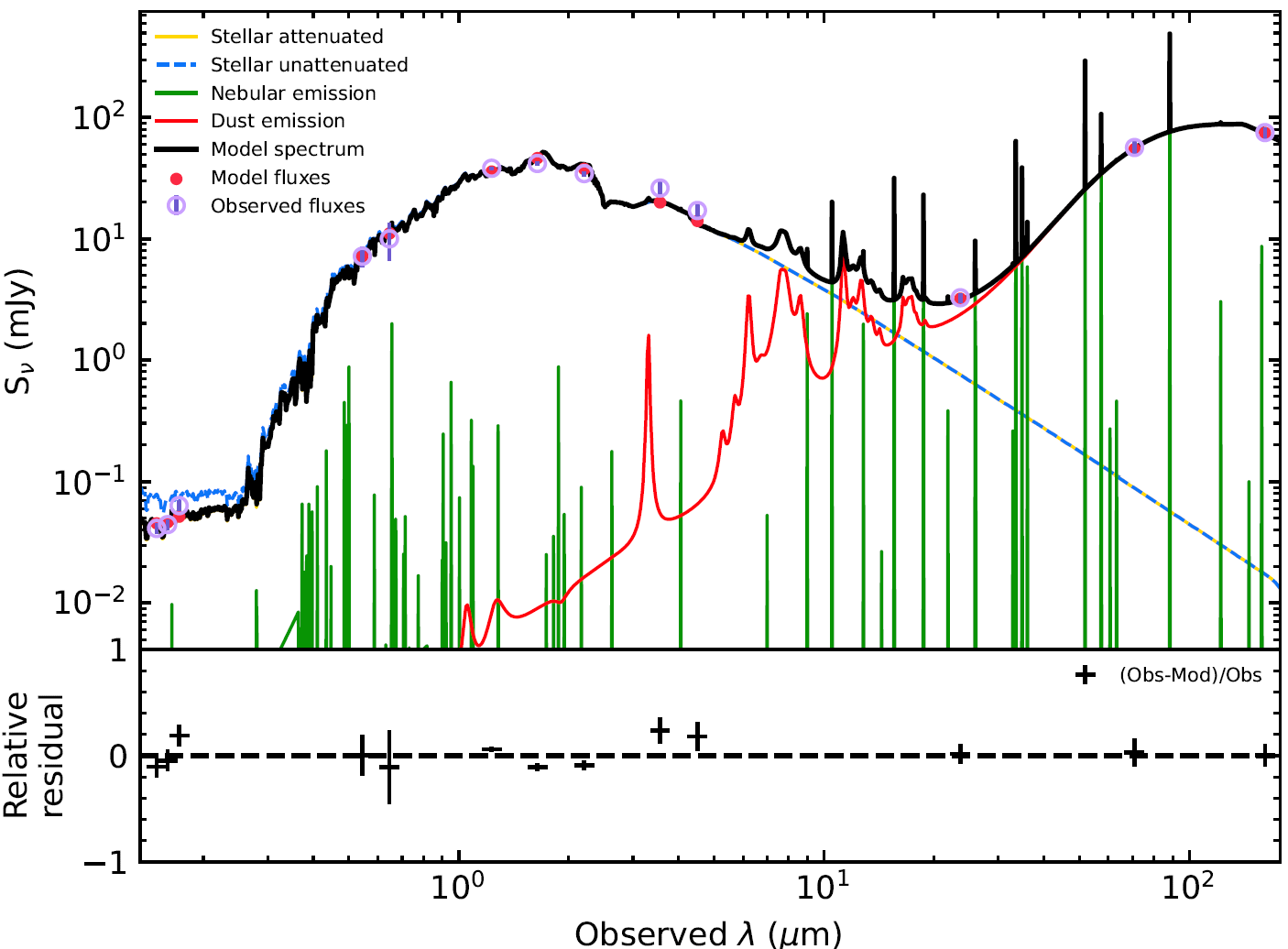}   \includegraphics[scale=0.48]{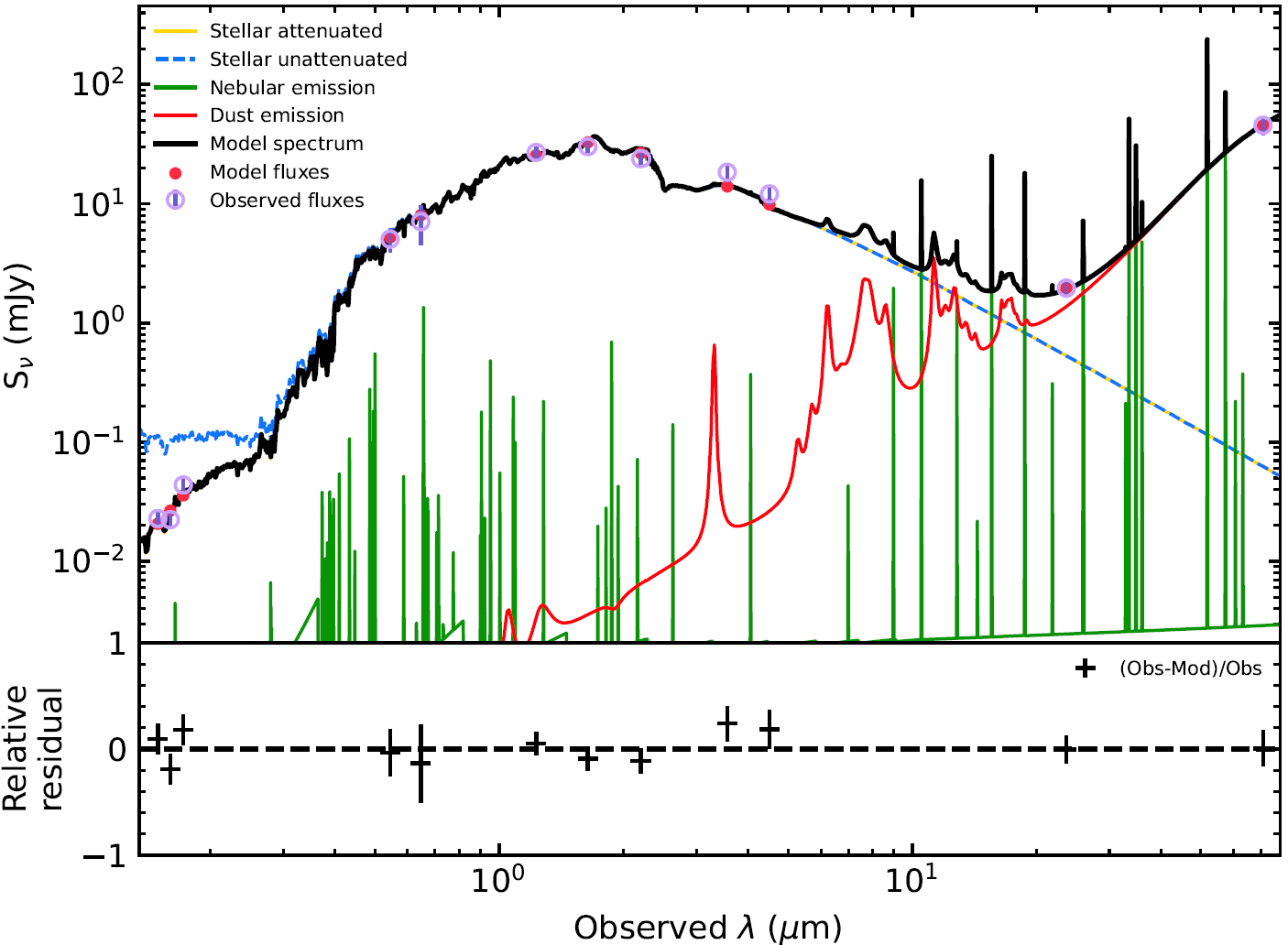} 
\includegraphics[scale=0.48]{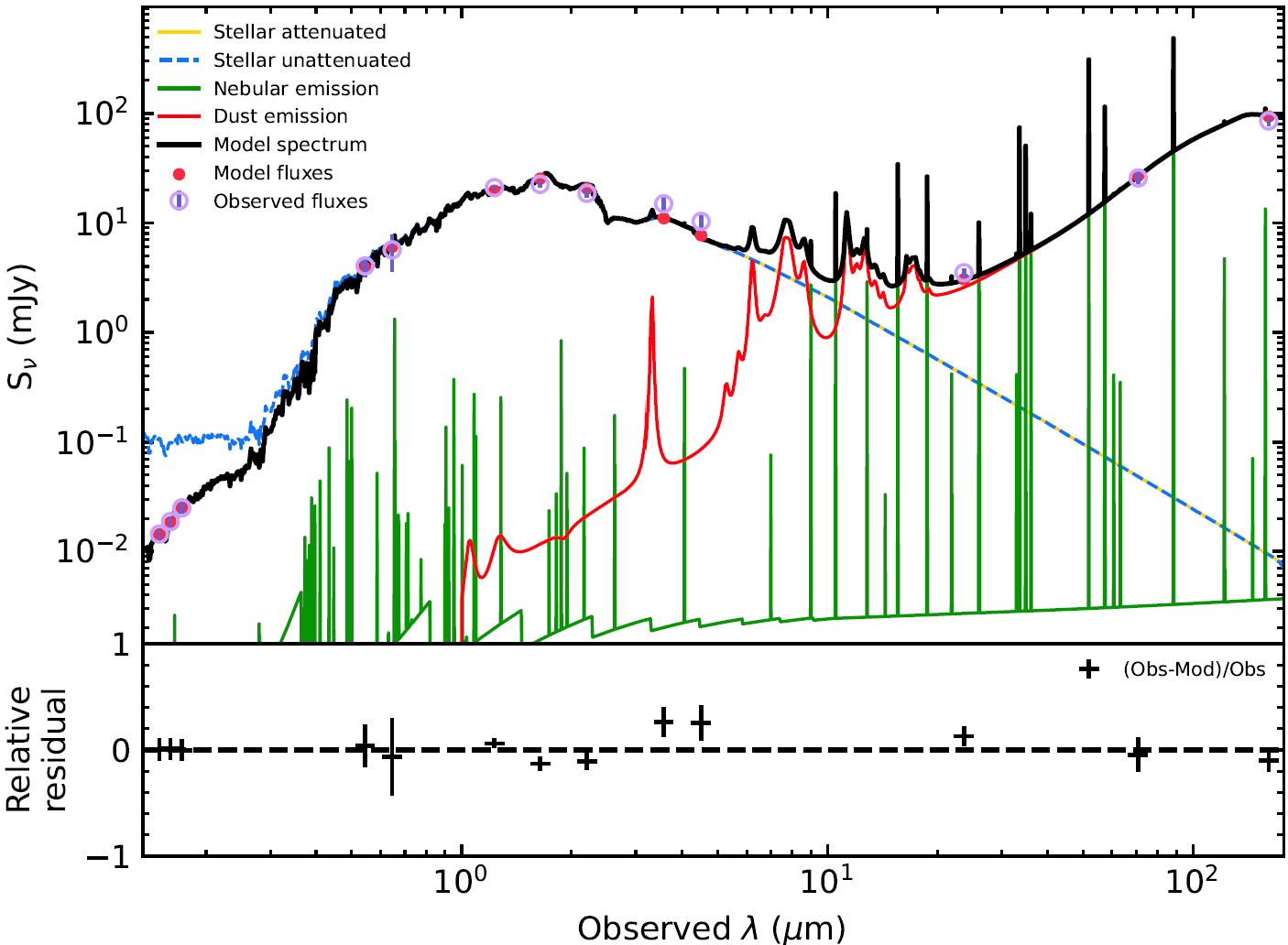}   \includegraphics[scale=0.48]{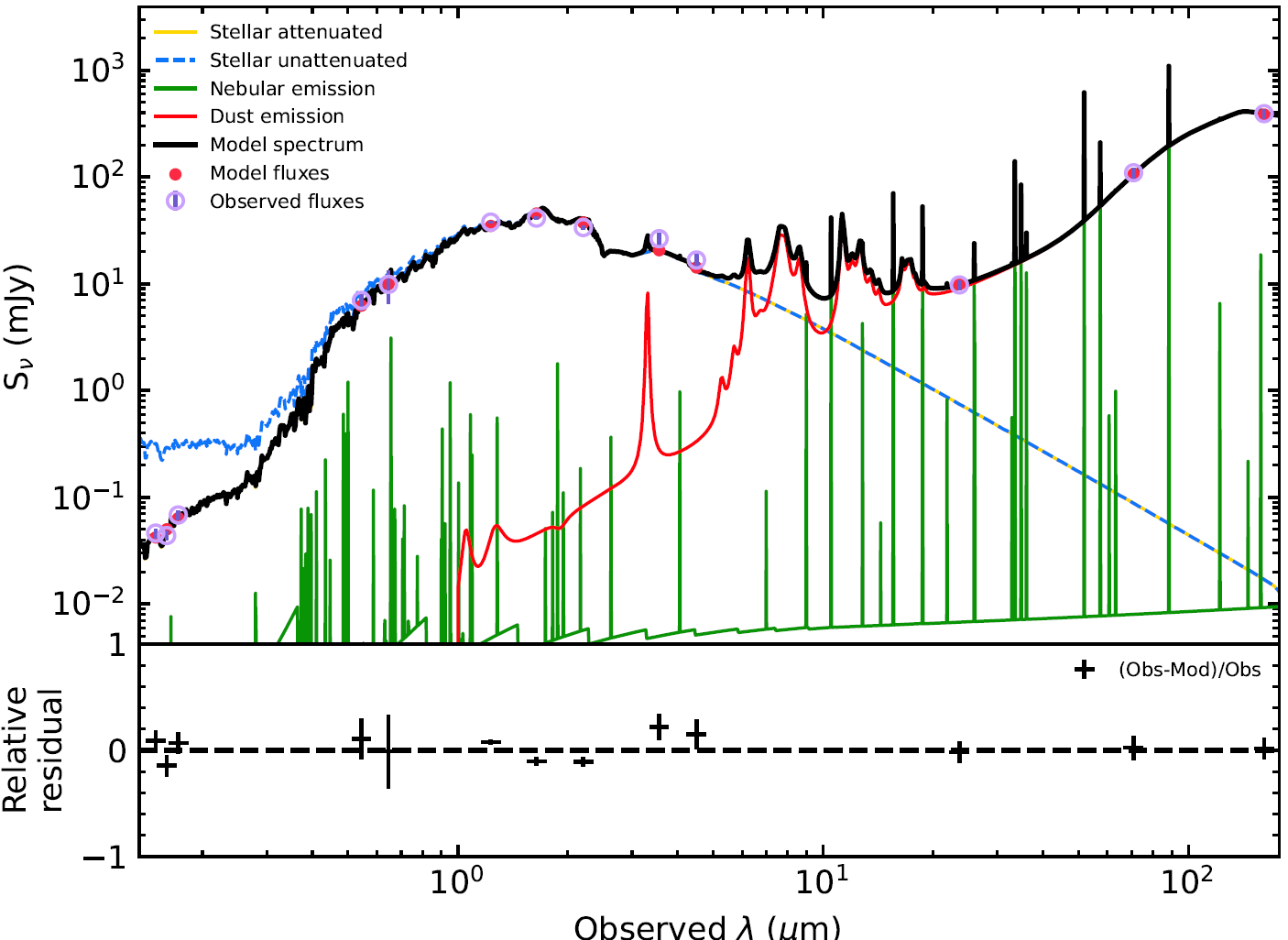} 
    	    \caption{Best-fit models and residual differences between model and data for the regions (in order of Region 2 to Region 28 from top left to bottom right). The fit for Region 1 is shown in Figure~\ref{fig:reg1fit}.
            The legend indicates symbols for input data, model points, and the contributions from the different components of the model flux density. Figure continues below.\label{fig:SEDfits}}
\end{figure*}

\begin{figure*}[ht]
    \centering
\includegraphics[scale=0.48]{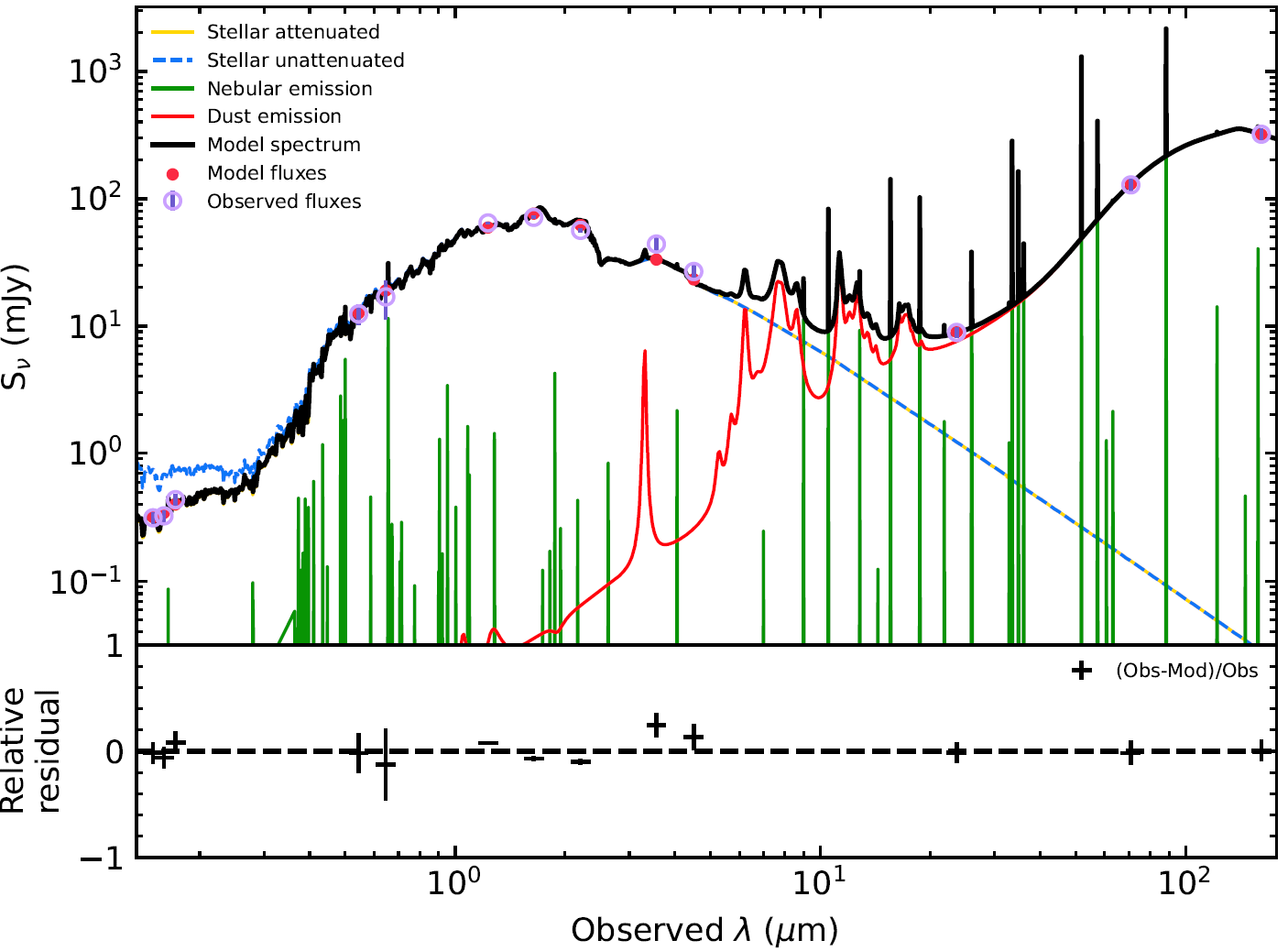}   \includegraphics[scale=0.48]{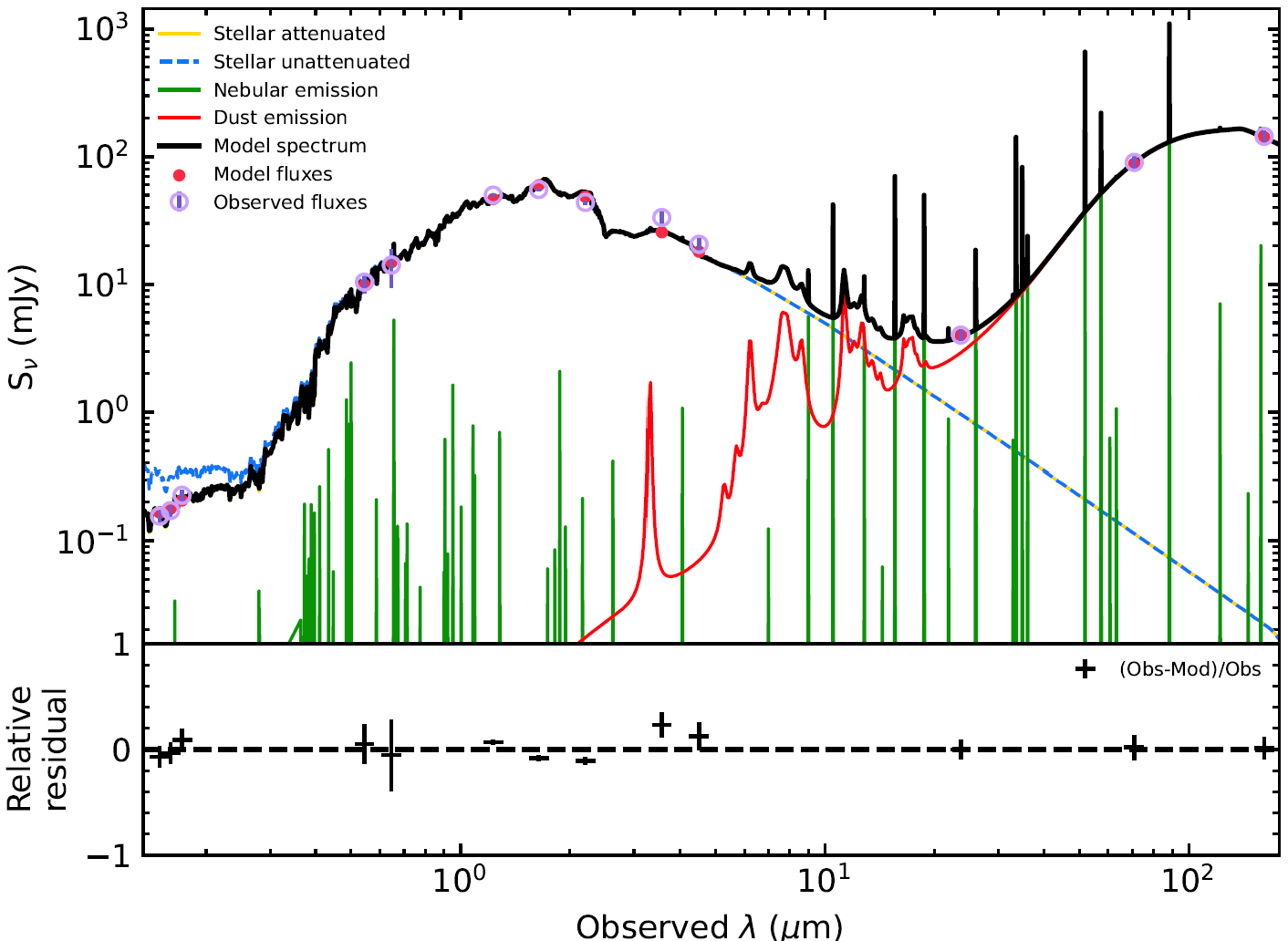} 
\includegraphics[scale=0.48]{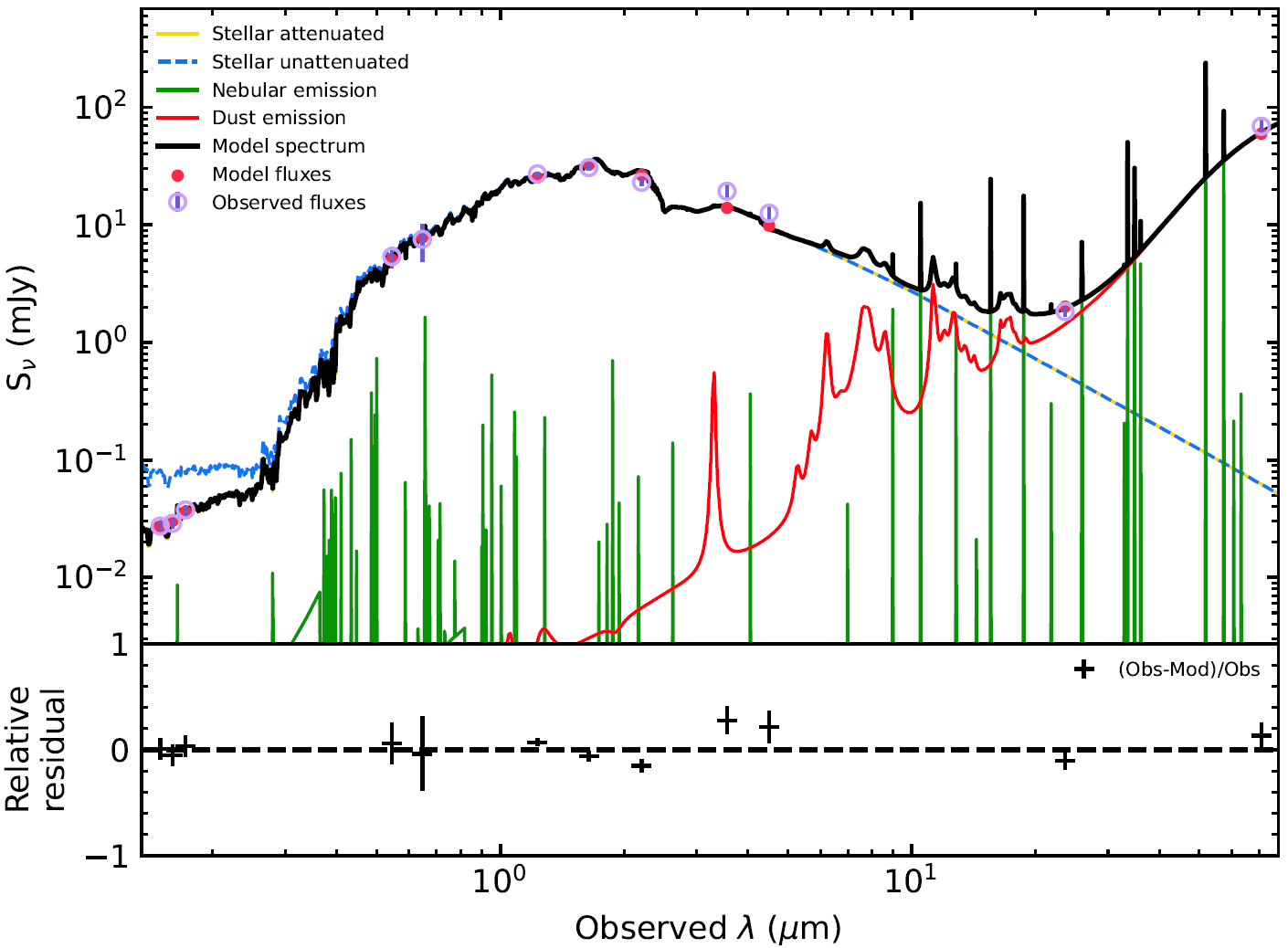}  \includegraphics[scale=0.48]{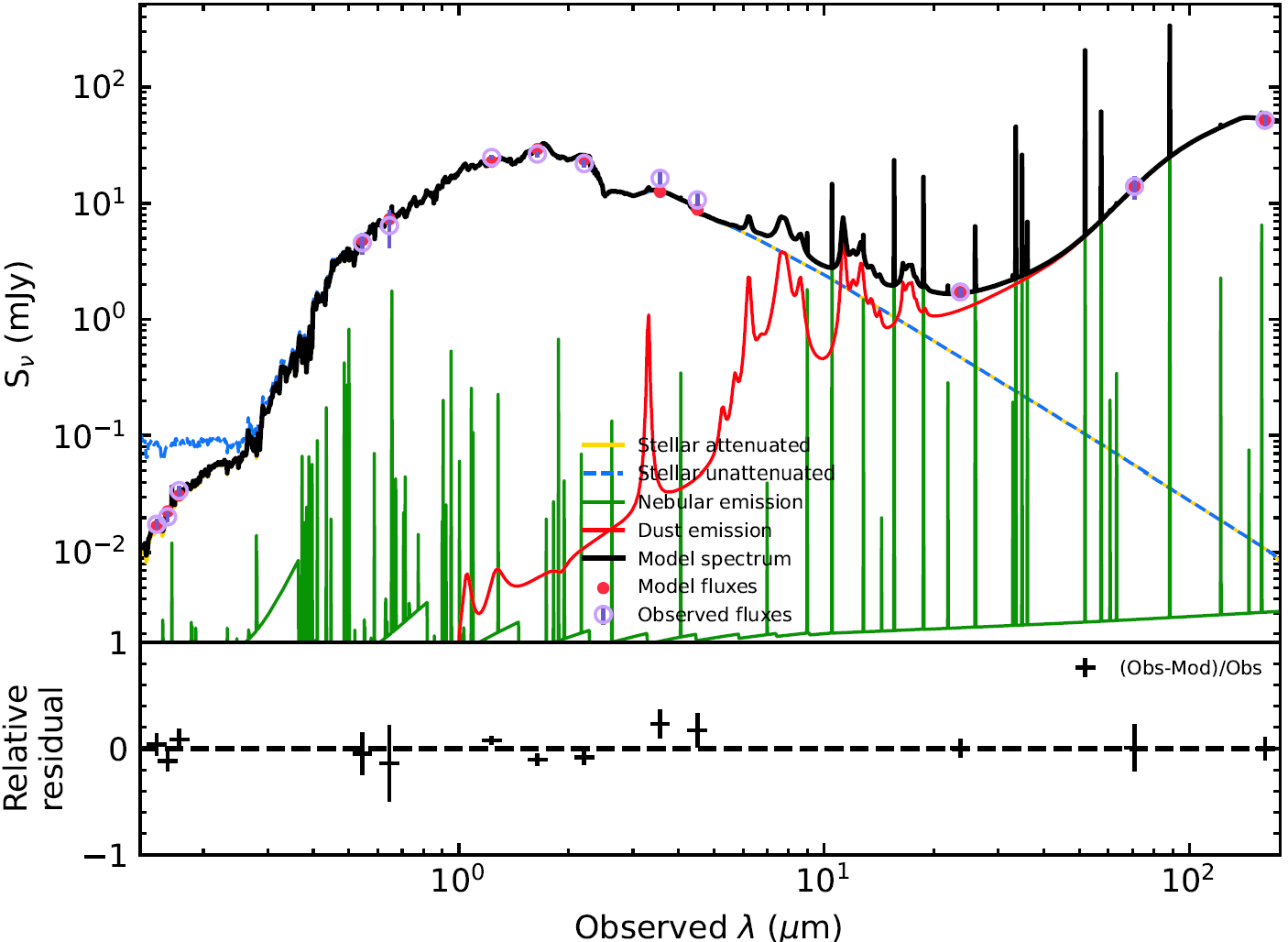} 
\includegraphics[scale=0.48]{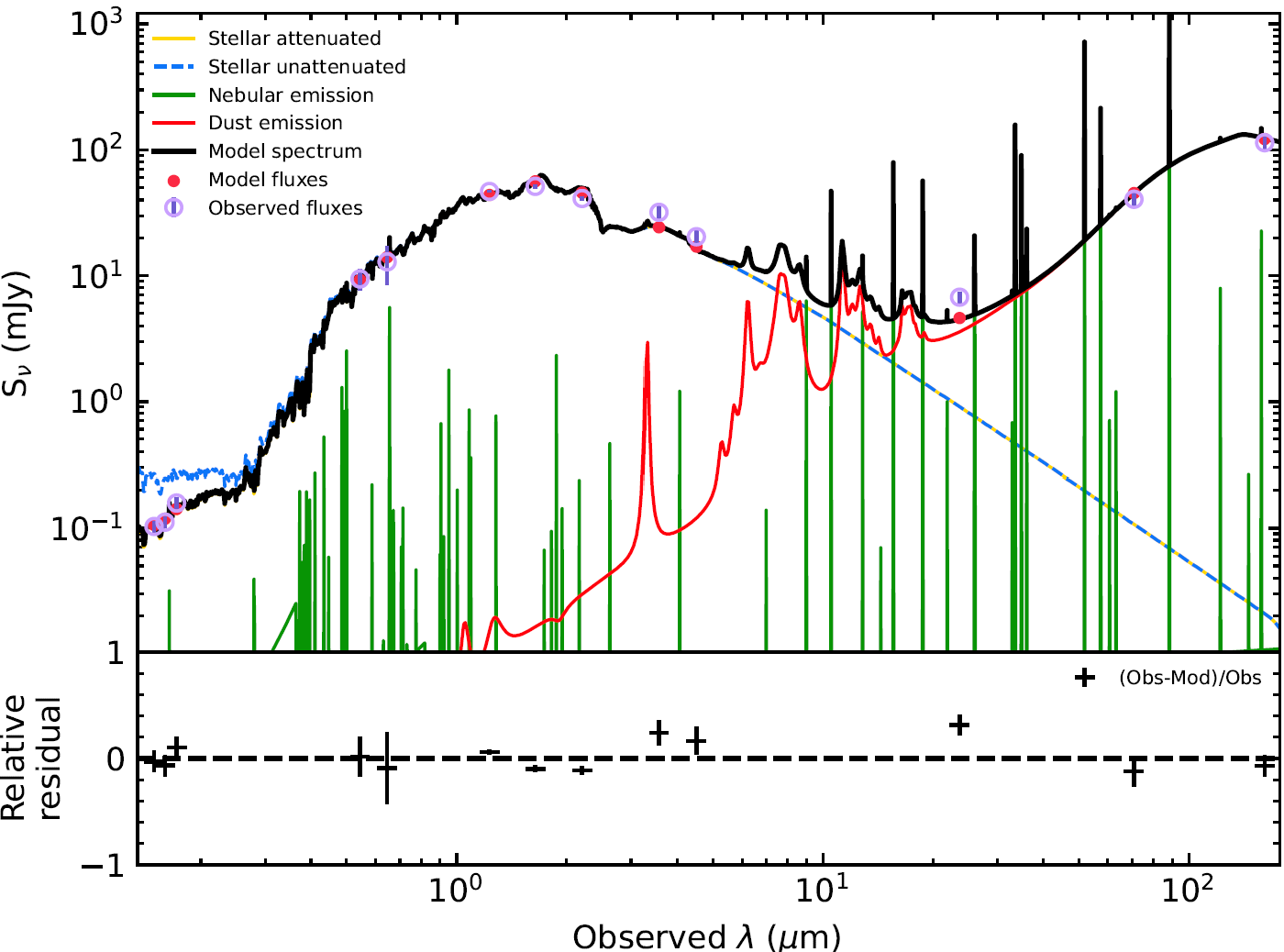}  \includegraphics[scale=0.48]{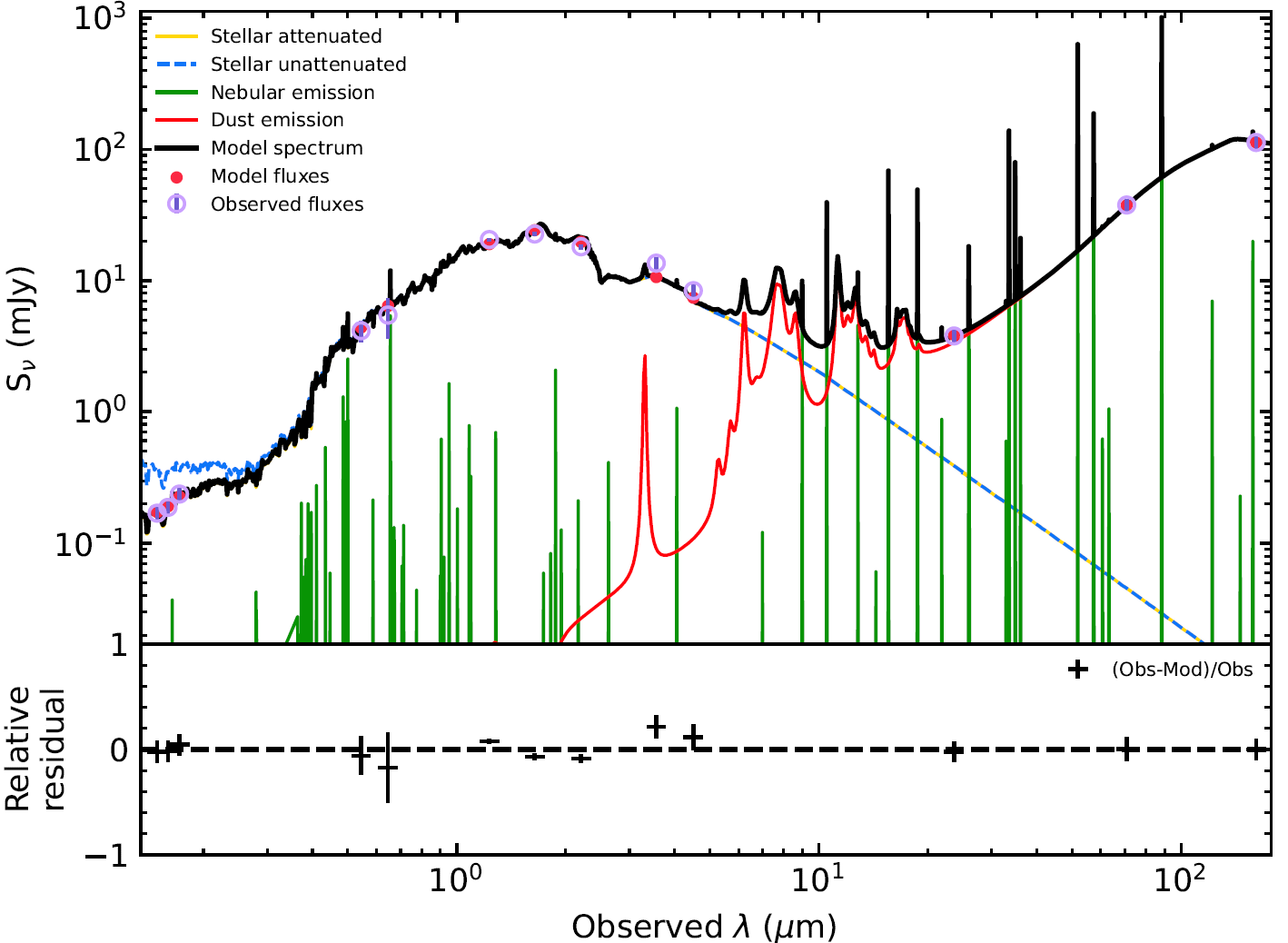} 
     \caption{Continued.}
\end{figure*}

\begin{figure*}[ht]
    \centering
\includegraphics[scale=0.48]{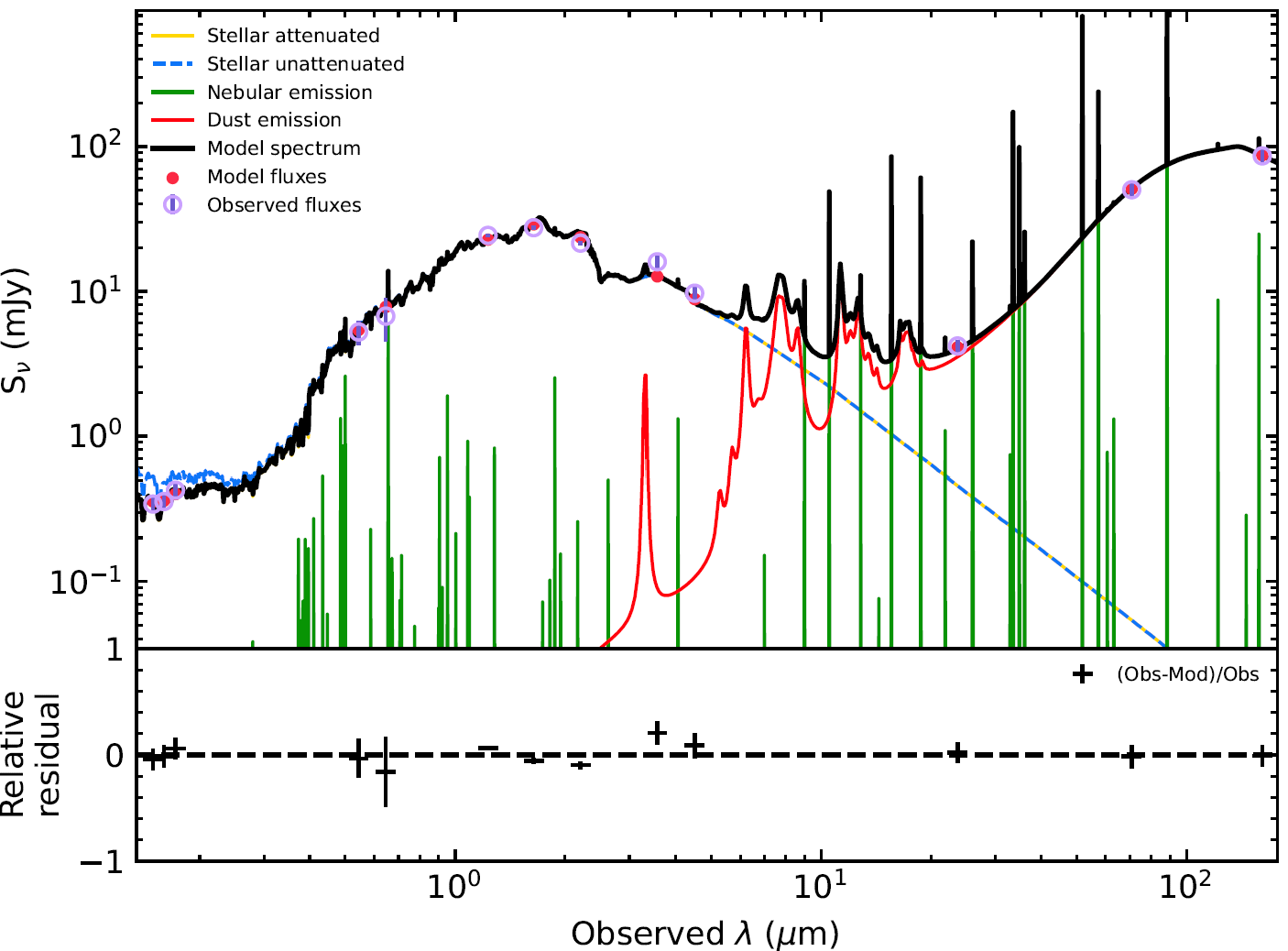}  \includegraphics[scale=0.48]{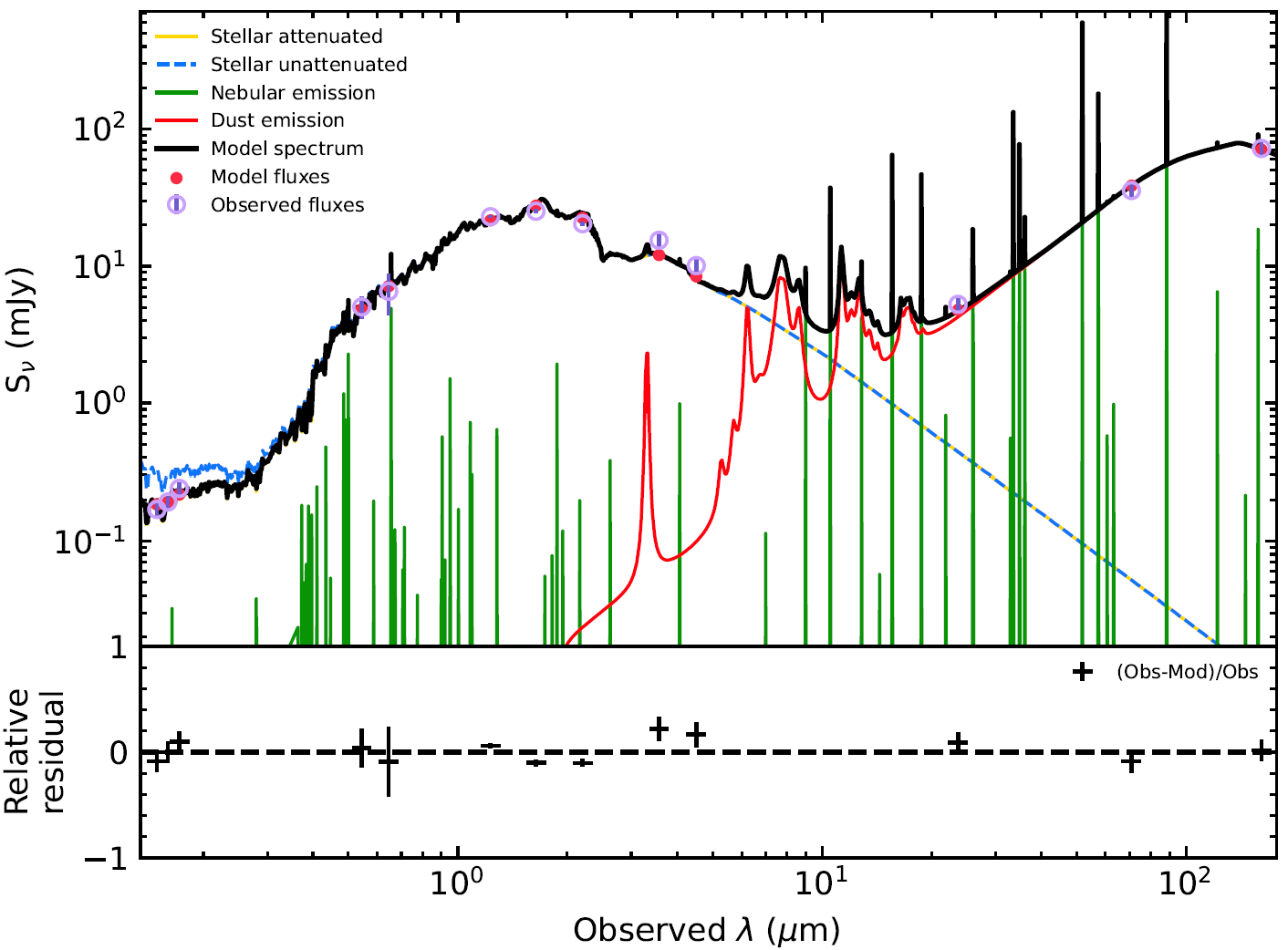} 
\includegraphics[scale=0.48]{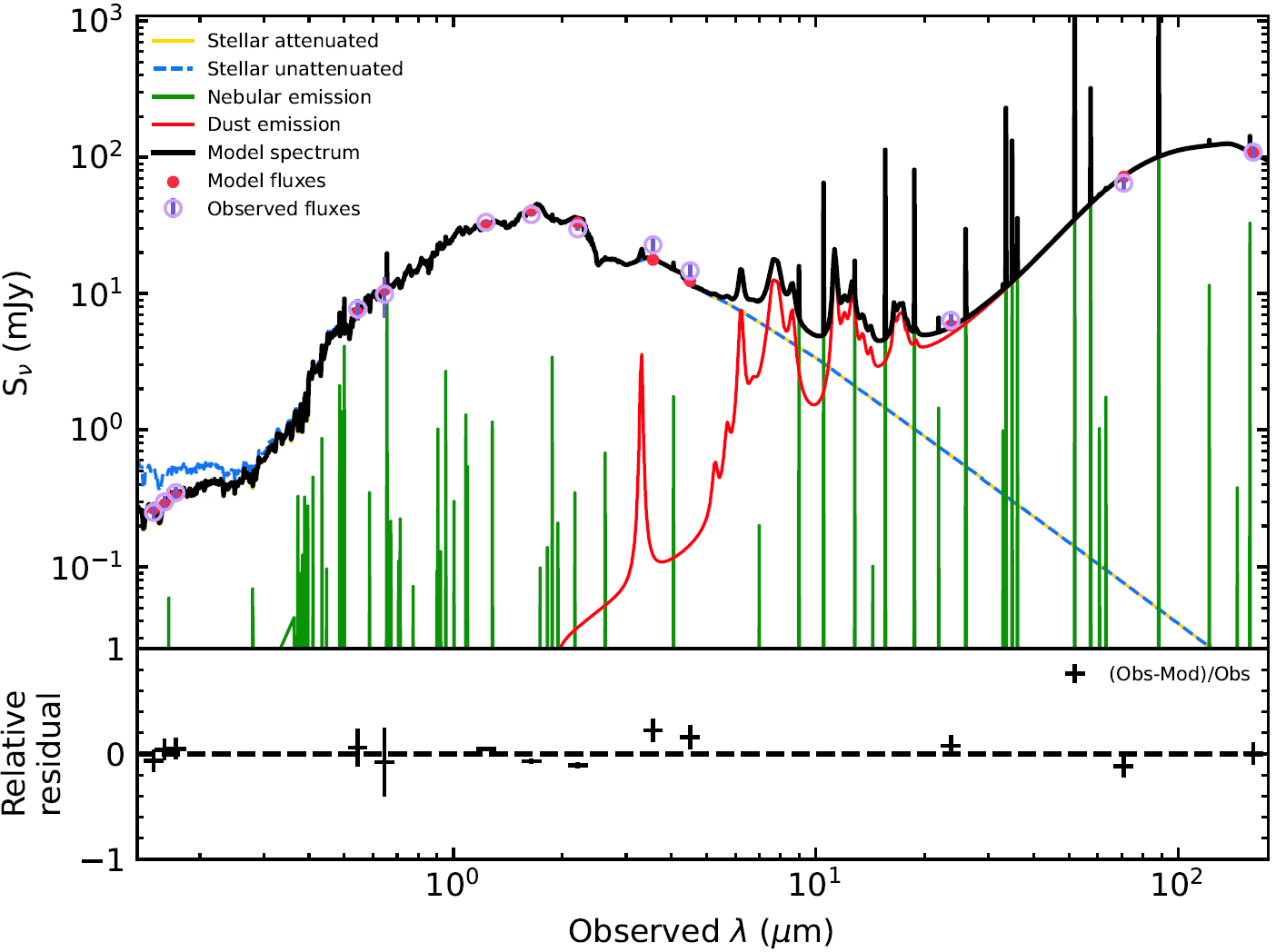}  \includegraphics[scale=0.48]{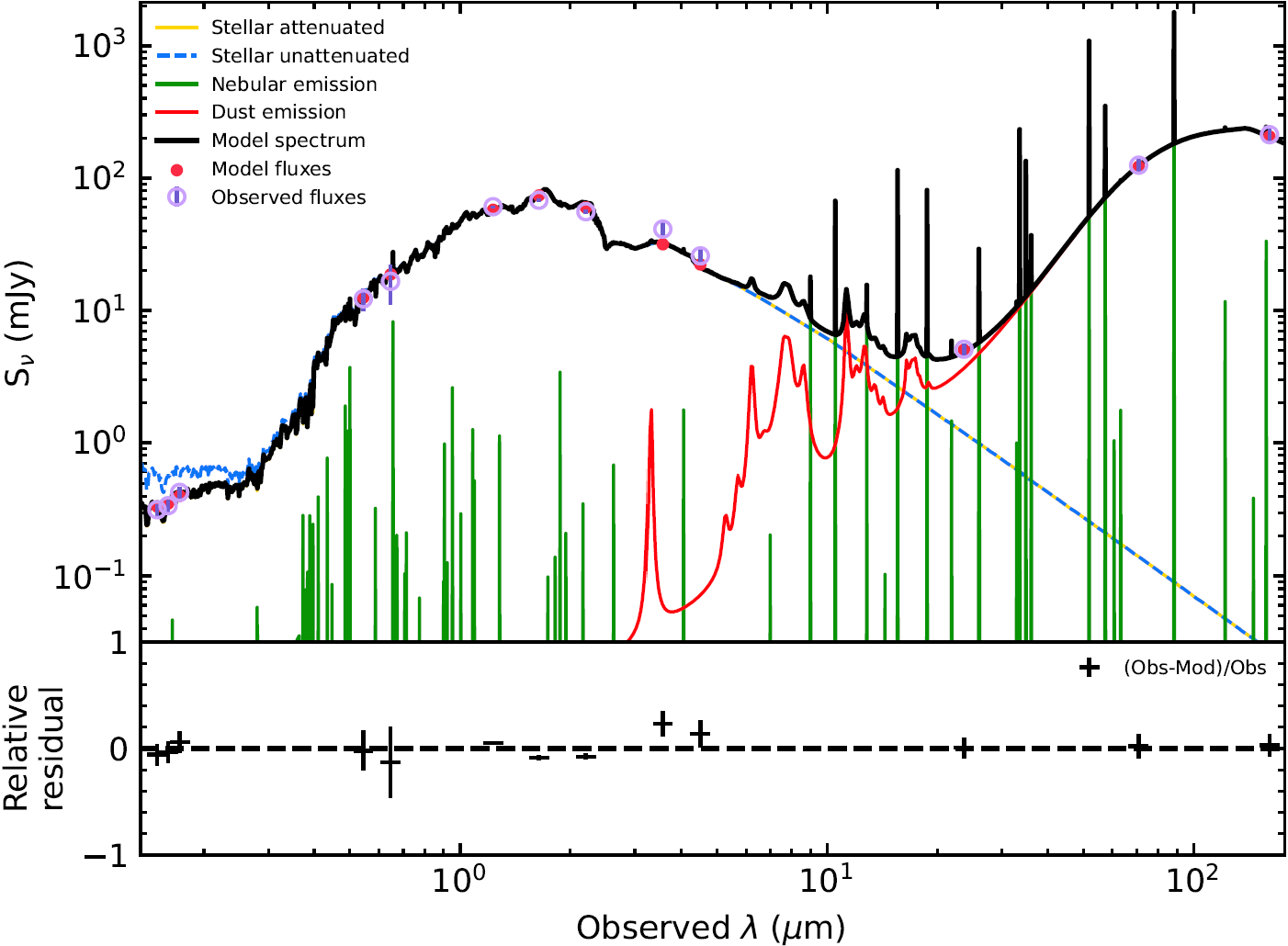} 
 \includegraphics[scale=0.47]{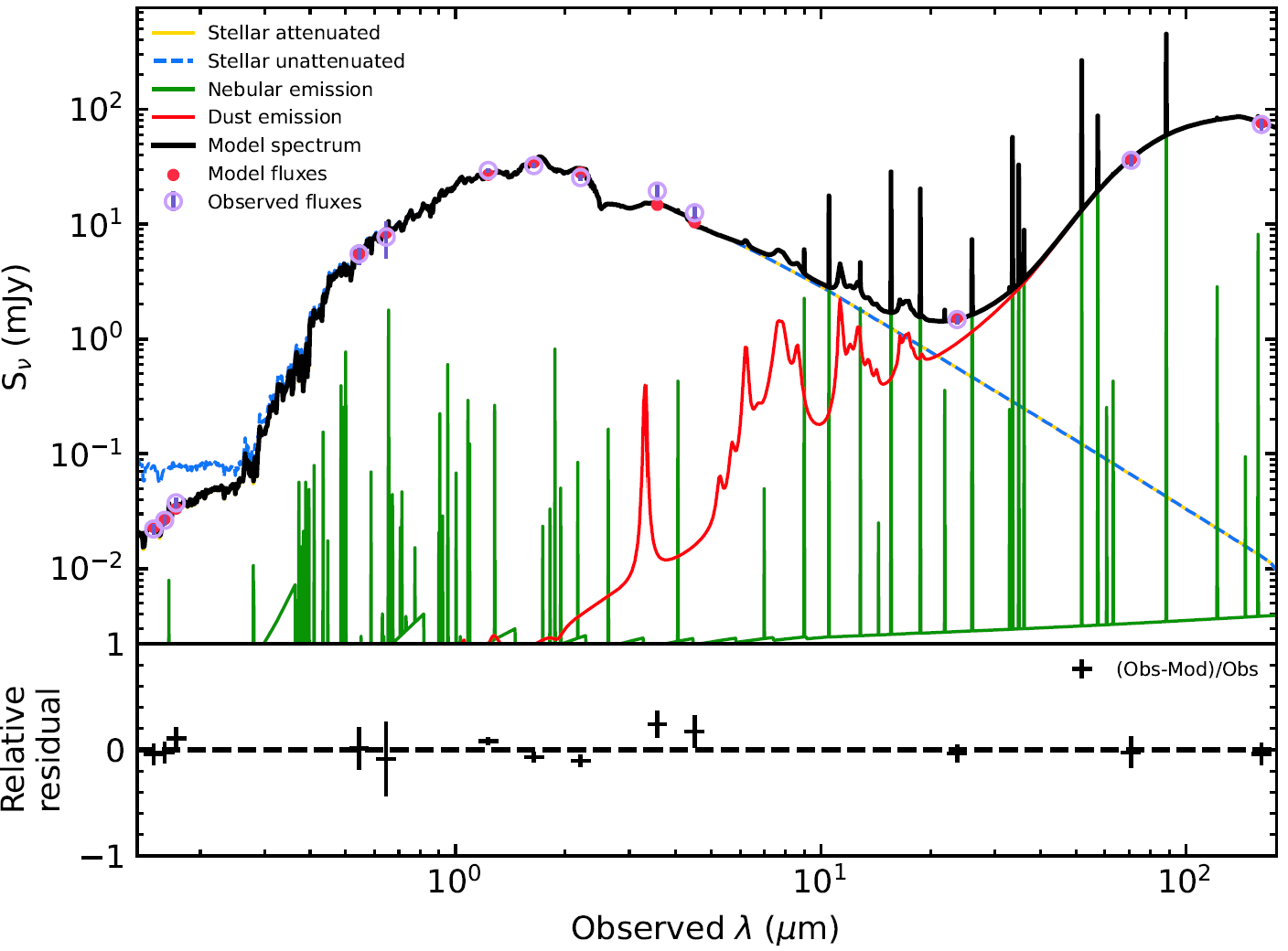} \includegraphics[scale=0.47]{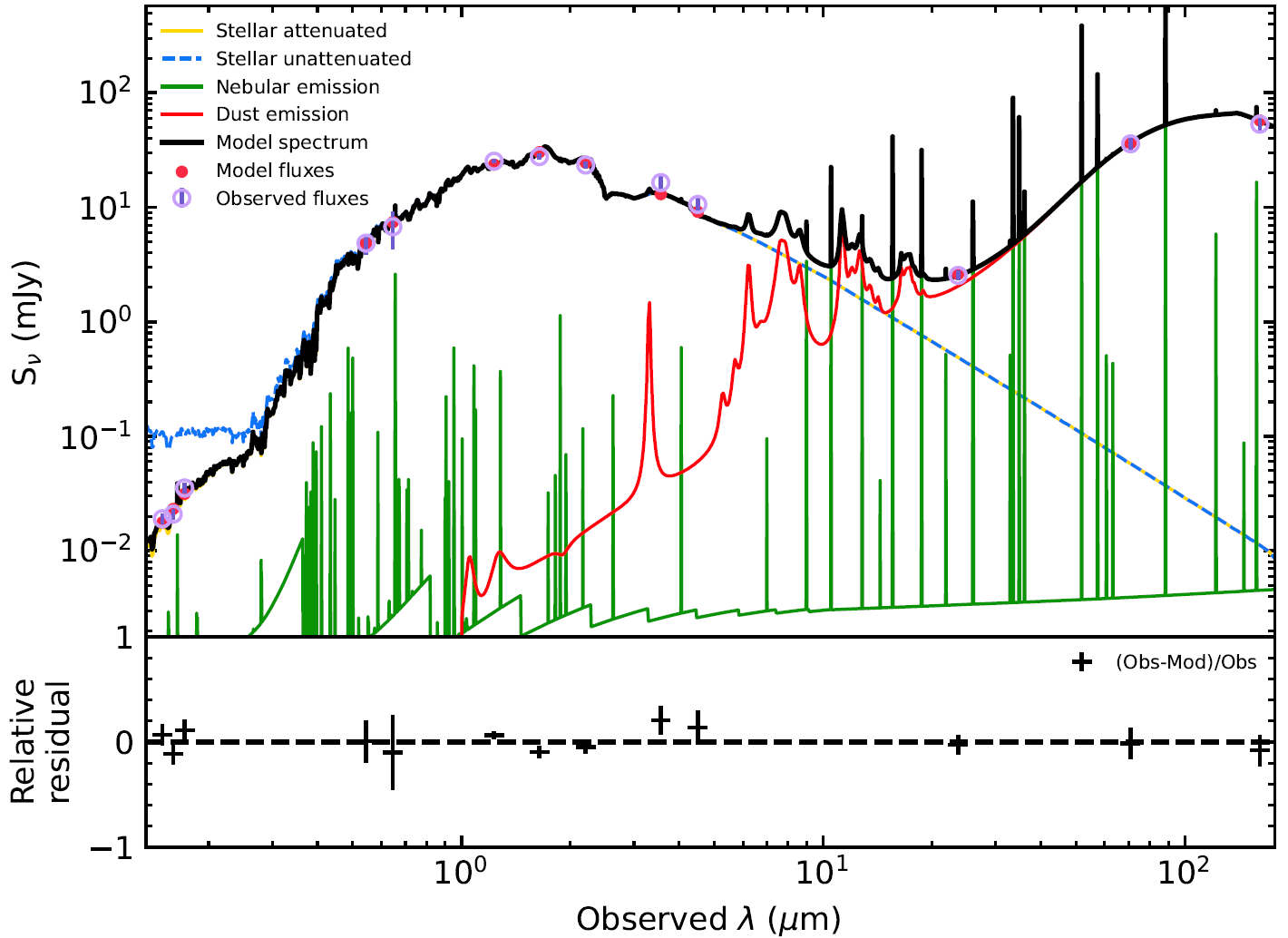} 
     \caption{Continued.}
\end{figure*}

 \begin{figure*}[ht]
    \centering
 \includegraphics[scale=0.47]{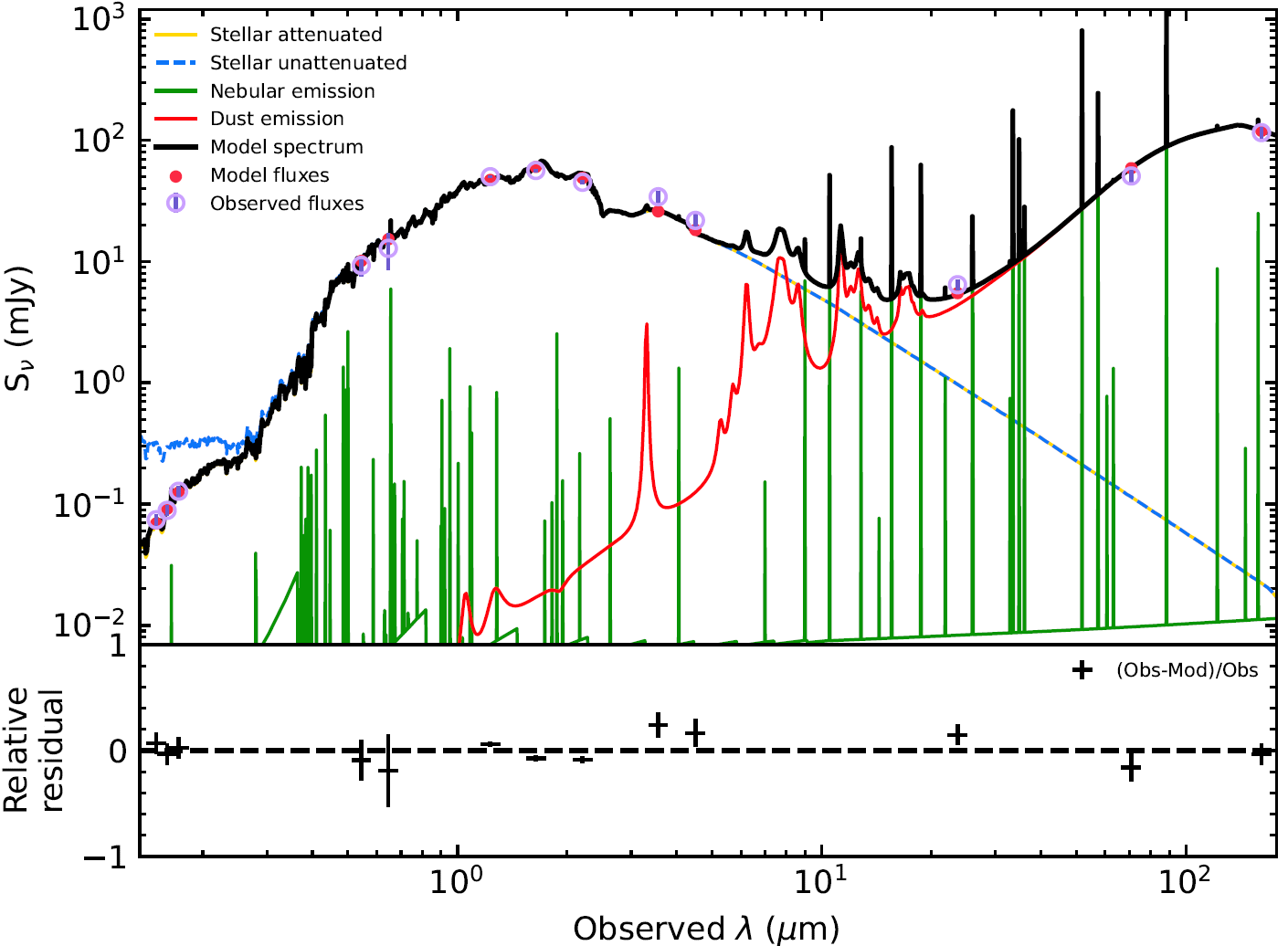} \includegraphics[scale=0.47]{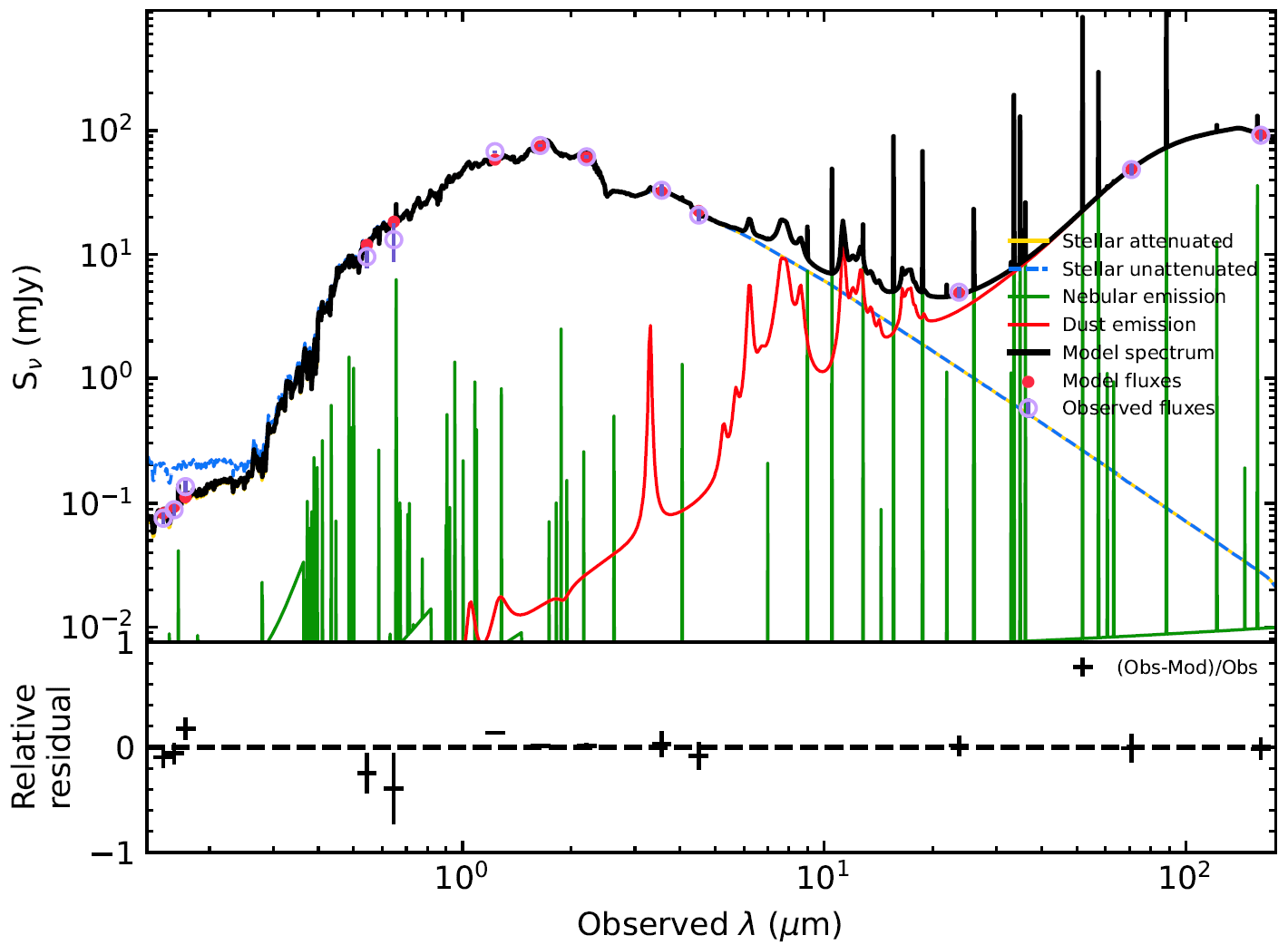} 
\includegraphics[scale=0.48]{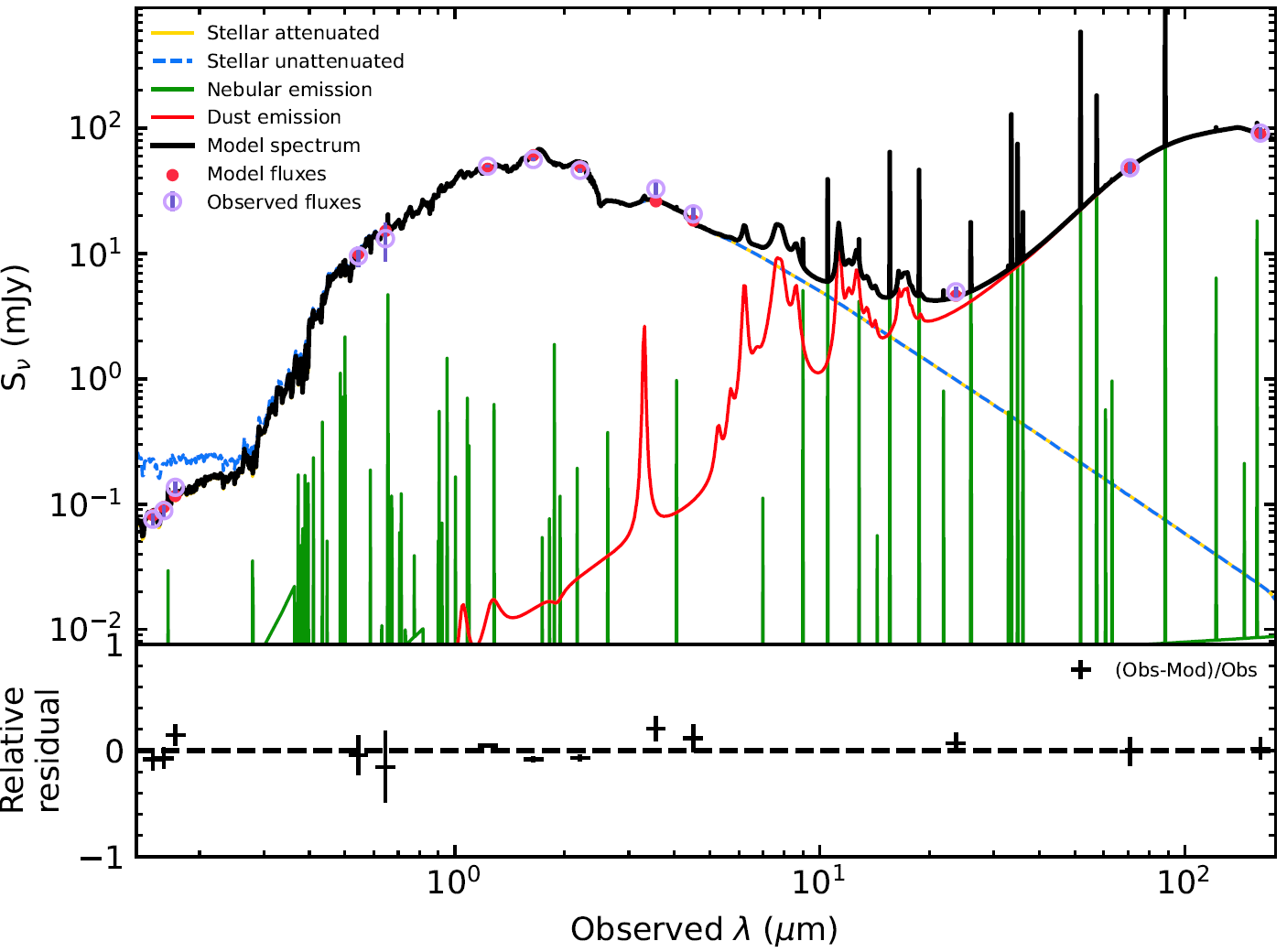}  \includegraphics[scale=0.48]{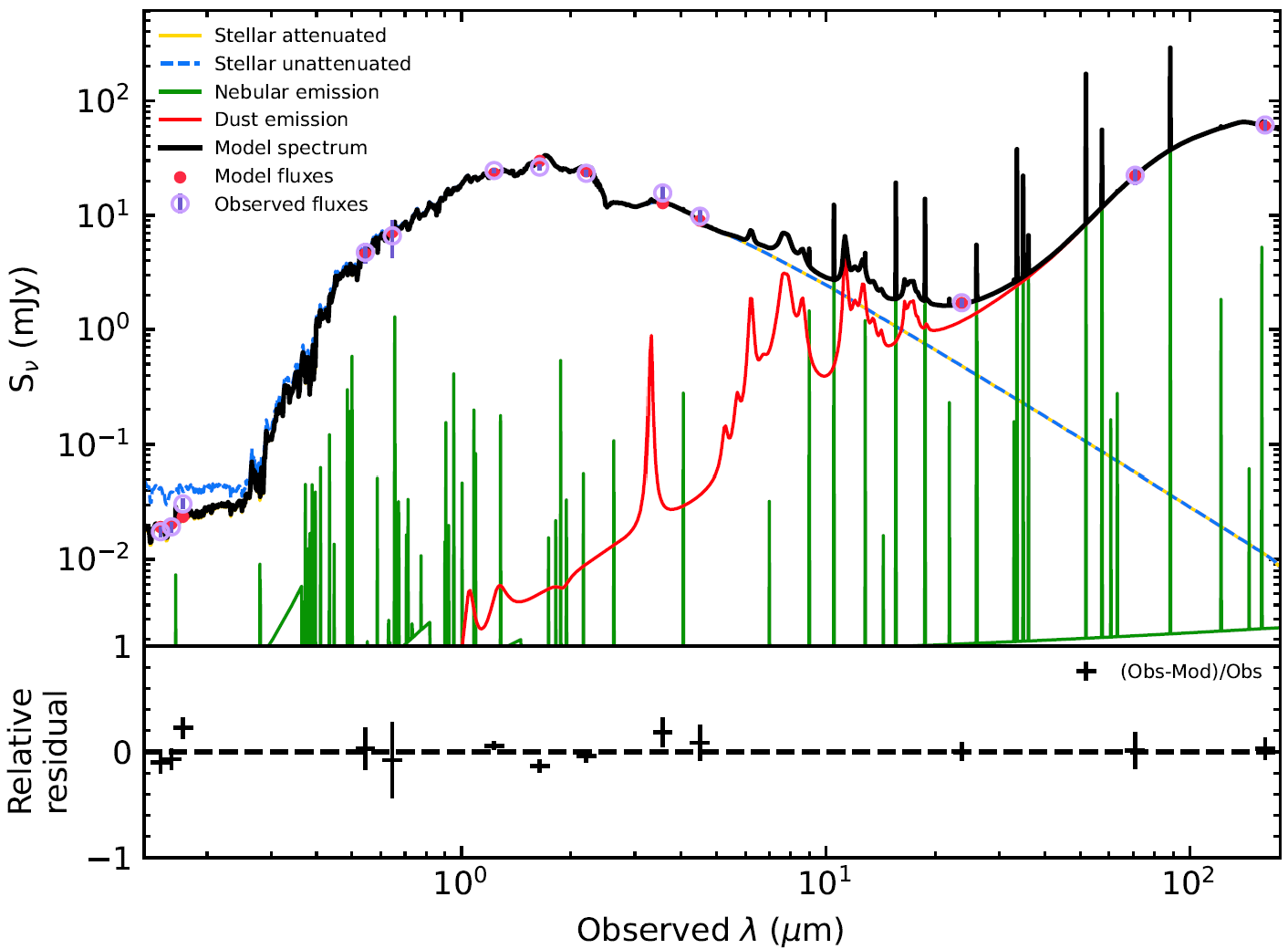} 
\includegraphics[scale=0.47]{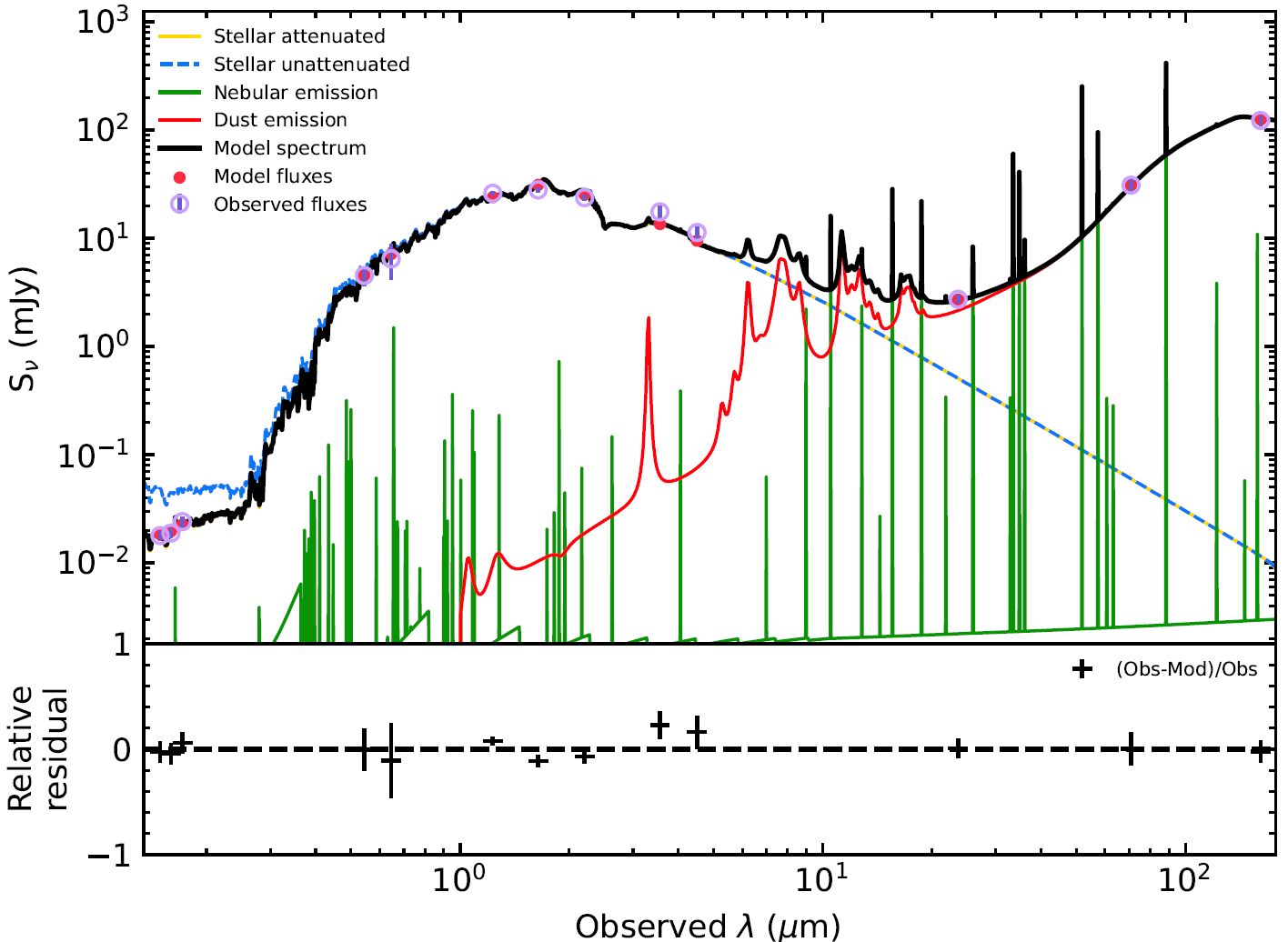} \includegraphics[scale=0.47]{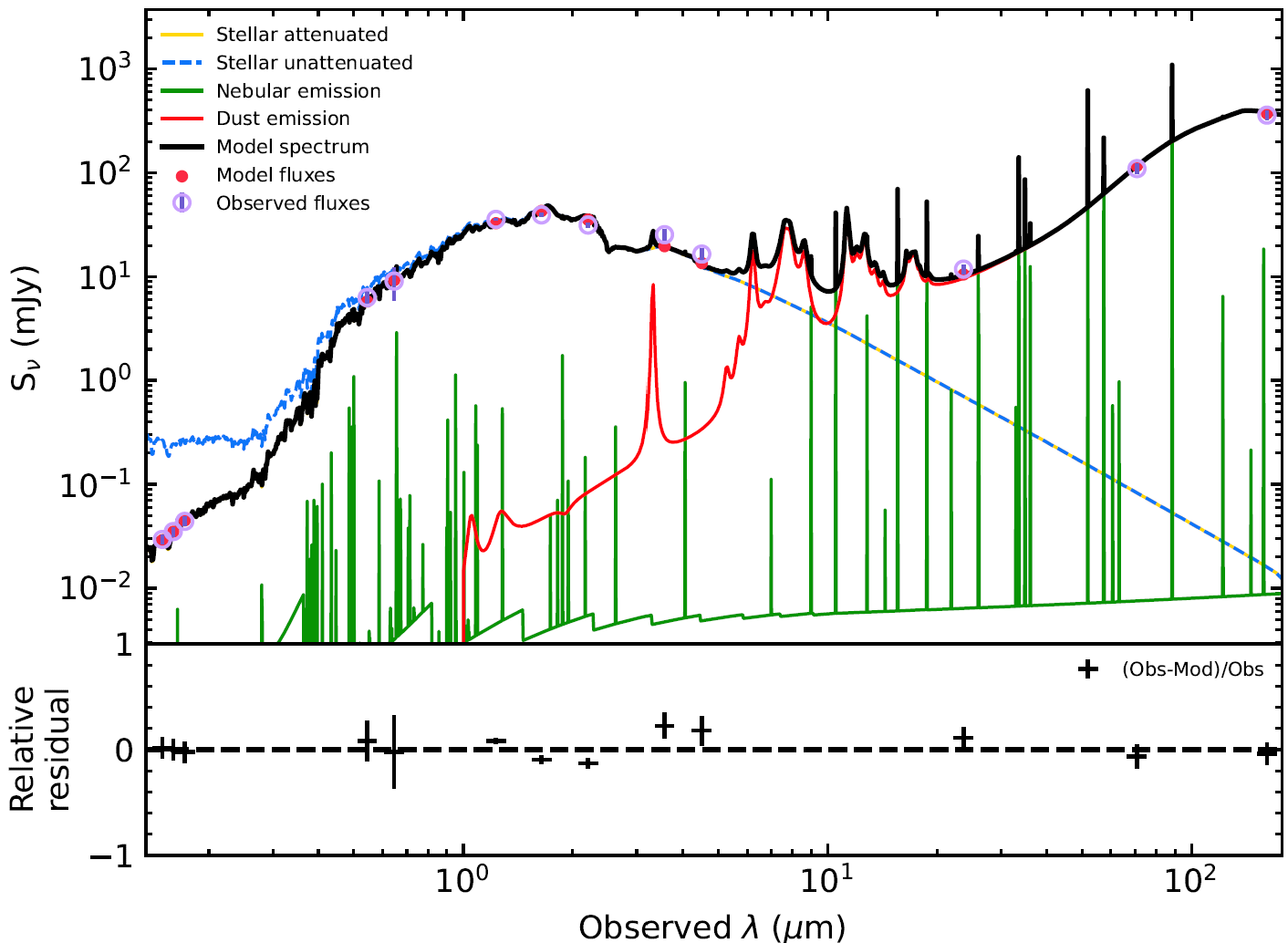}
      \caption{Continued.}
\end{figure*}

\begin{figure*}[ht]
    \centering
 \includegraphics[scale=0.48]{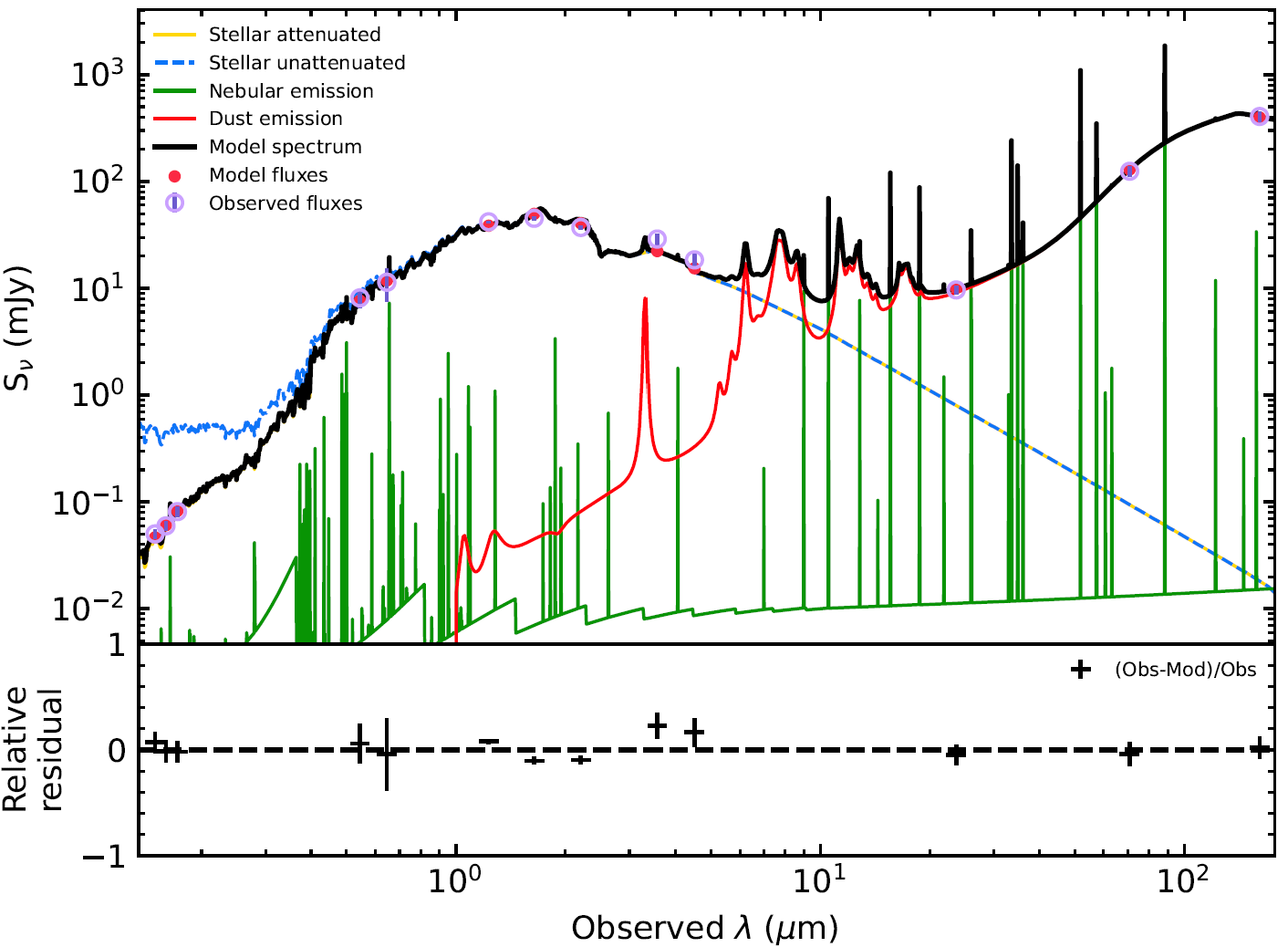}  \includegraphics[scale=0.48]{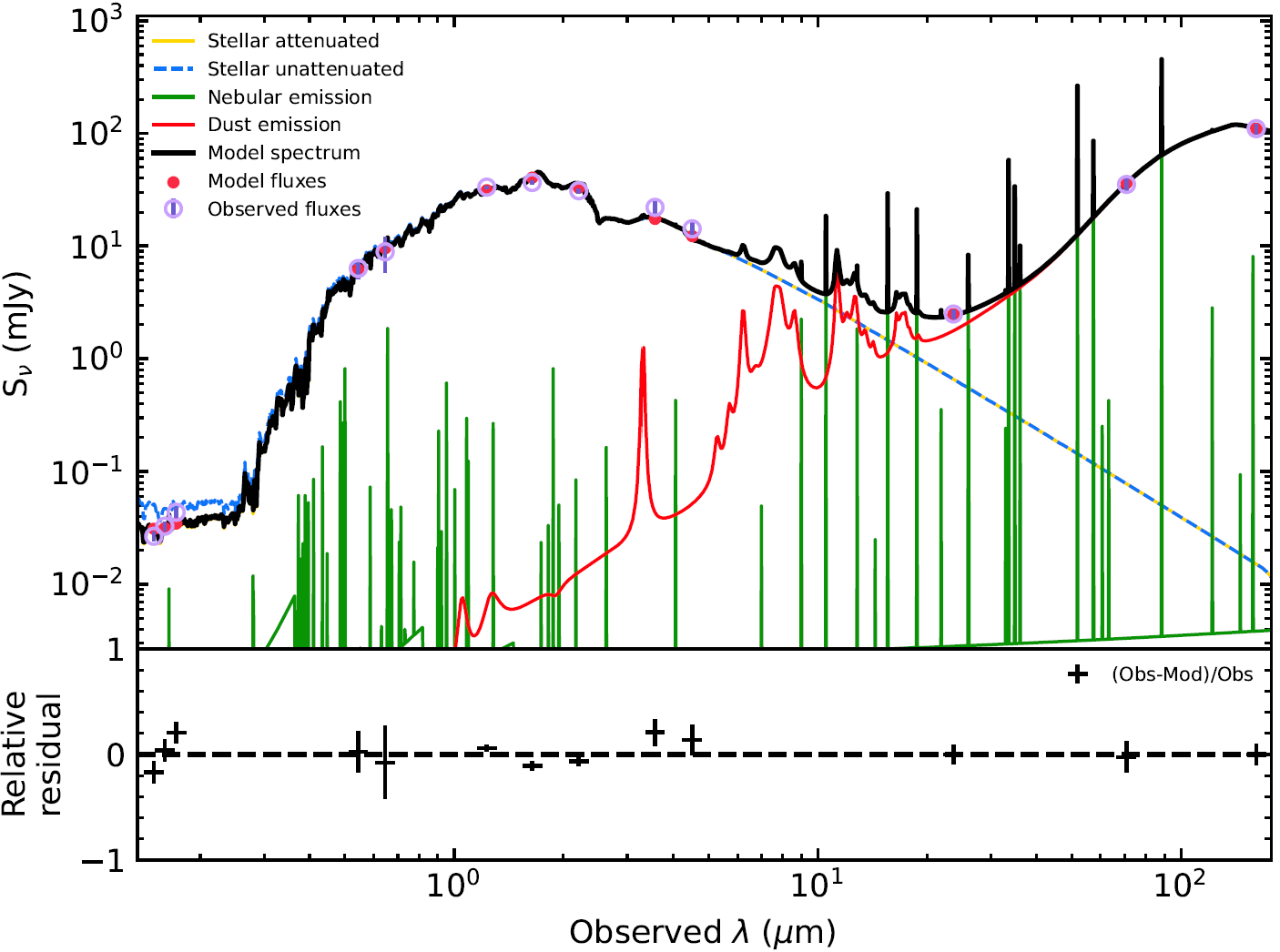}  \includegraphics[scale=0.48]{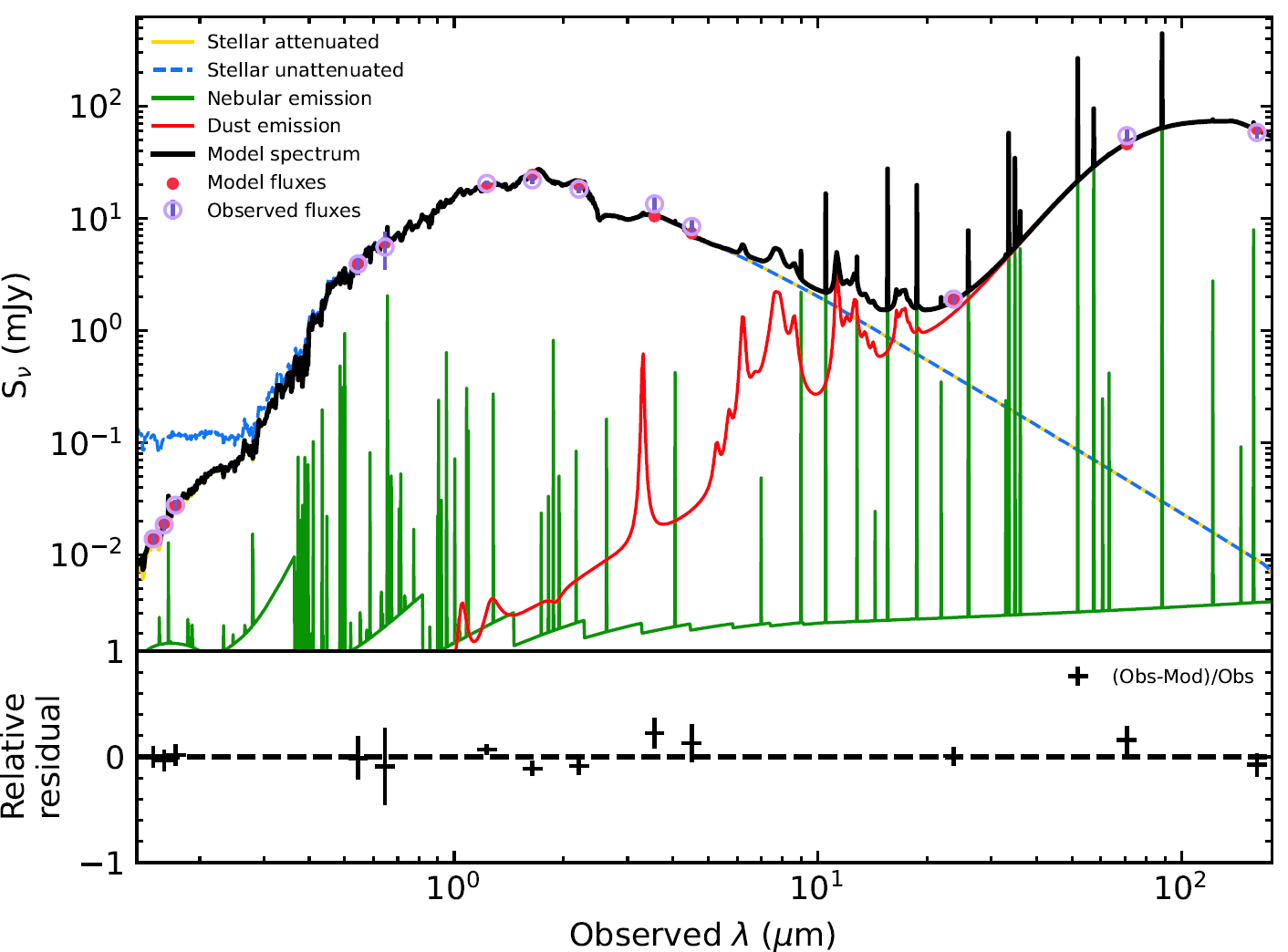}
     \caption{Continued.}
\end{figure*}

\end{document}